\documentclass[prd,nofootinbib,amsfonts,tightenlines,singlecolumn]{revtex4-1}
\usepackage{graphicx} 
\usepackage{amsmath}
\usepackage{amssymb}
\usepackage{ae} 
\usepackage{aecompl}
\usepackage{bm} 
\usepackage[usenames,dvipsnames]{color}
\usepackage[pdfa, bookmarks=true,bookmarksnumbered=true]{hyperref}
 \hypersetup{colorlinks=true,
     linkcolor=Blue,          % color of internal links (change box color 
     citecolor=Mahogany,
     urlcolor=Mahogany
     }

\usepackage{amsmath}
\usepackage{amssymb}
\usepackage{epsfig}
\usepackage{fontenc}
\usepackage{graphicx}
\usepackage[lofdepth,lotdepth,caption=false]{subfig}
\usepackage{color}
\usepackage{booktabs}
\usepackage{tabularx}
\usepackage{multirow}
\usepackage{longtable}
\usepackage[normalem]{ulem}
\usepackage{dcolumn}
\usepackage{rotating}%
\usepackage{hyperref}

\newcommand{\bi}{\begin{itemize}}
\newcommand{\ei}{\end{itemize}}

\def\beq{\begin{equation}}
\def\eeq{\end{equation}}

\newcommand{\bea}{\begin{eqnarray}}
\newcommand{\eea}{\end{eqnarray}}

\newcommand{\pmm}{P_{\mu \mu}}
\newcommand{\pme}{P_{\mu e}}

\newcommand{\ldm}{\Delta m_{31}^2}

\newcommand{\ie}{{\it i.e.}}

\newcommand{\eg}{{\it e.g.}}

\newcommand{\eet}{\varepsilon_{e\tau}}
\newcommand{\emt}{\varepsilon_{\mu\tau}}
\newcommand{\ett}{\varepsilon_{\tau\tau}}
\newcommand{\eee}{\varepsilon_{ee}}
\newcommand{\eem}{\varepsilon_{e\mu}}

\newcommand{\eeta}{|\varepsilon_{e\tau}|}
\newcommand{\emta}{|\varepsilon_{\mu\tau}|}

\newcommand{\eema}{|\varepsilon_{e\mu}|}

\newcommand{\eetp}{\varphi_{e\tau}}
\newcommand{\emtp}{\varphi_{\mu\tau}}
\newcommand{\eemp}{\varphi_{e\mu}}
\def\epsilon{\varepsilon}

\newcommand\sch{Schr$\ddot{\rm o}$dinger~}

\newcommand{\dune}{{\sc DUNE}}

\def\<{\langle}
\def\>{\rangle}

\def\dfrac#1#2{{\displaystyle\frac{#1}{#2}}}

\def\lsim{\mathrel{\rlap{\lower4pt\hbox{\hskip1pt$\sim$}}
    \raise1pt\hbox{$<$}}}         %less than or approx. symbol
\def\gsim{\mathrel{\rlap{\lower4pt\hbox{\hskip1pt$\sim$}}
    \raise1pt\hbox{$>$}}}         %greater than or approx. symbol

\begin{document}  

\title{Correlations and degeneracies among the NSI parameters with tunable beams at DUNE 
}

\author{Mehedi Masud}
\email{masud@ific.uv.es}
\affiliation{Astroparticle and High Energy Physics Group, Instituto de F\'{i}sica Corpuscular (CSIC/Universitat de Val\`{e}ncia), Parc Cientific de Paterna. 
  C/Catedratico Jos\'e Beltr\'an, 2 E-46980 Paterna (Val\`{e}ncia) - Spain}

\author{Samiran Roy}
\email{samiranroy@hri.res.in}
\affiliation{Harish-Chandra Research Institute, Chattnag Road, Allahabad 211 019, India}
\affiliation{Homi Bhabha National Institute, Training School Complex, Anushakti Nagar,
     Mumbai 400085, India}

\author{Poonam Mehta}
\email{pm@jnu.ac.in}
\affiliation{School of Physical Sciences, Jawaharlal Nehru University, 
      New Delhi 110067, India}

\date{\today}
%===========================================================================
\begin{abstract}

The Deep Underground Neutrino Experiment (DUNE) is a leading experiment in neutrino physics which is presently under construction. DUNE aims to measure the yet unknown parameters in the three flavor oscillation scenario which includes discovery of leptonic CP violation, determination of the mass hierarchy and determination of the octant of $\theta_{23}$. Additionally, the ancillary goals of DUNE include probing the subdominant effects induced by new physics. A widely studied new physics scenario is that of nonstandard neutrino interactions (NSI) in propagation which impacts the oscillations of neutrinos.~We consider some of the essential NSI parameters impacting the oscillation signals  at DUNE and explore the space of NSI parameters as well as study their correlations among themselves and with the yet unknown CP violating phase, $\delta$ appearing in the standard paradigm. The experiment utilizes a wide band beam and provides us with a unique opportunity to utilize different beam tunes at DUNE. We demonstrate that combining information from different beam tunes (low energy and medium energy) available at DUNE impacts the ability to probe some of these parameters and leads to altering the allowed regions in two-dimensional space of parameters considered. 

\end{abstract}
\maketitle
%
%----------------------------------------------------------------------%
\section{Introduction}

In a seminal paper in 1978, Wolfenstein first proposed the possibility that nonstandard neutrino interactions  (NSI) could be responsible for  conversion of a given neutrino flavour to another even if neutrinos were massless~\cite{Wolfenstein:1977ue}. However, thanks to the wealth of data accumulated by a variety of oscillation experiments covering different energies and baselines, we now have a  fairly clear picture that  neutrino oscillations occur due to nonzero neutrino masses. 
The data from most of the oscillation experiments can be nicely  explained by invoking three flavors of neutrinos ($\nu_e, \nu_\mu, \nu_\tau$) which are superpositions of the mass states $\nu_1,\nu_2,\nu_3$ with masses $m_1,m_2,m_3$ respectively. The $3\times 3$ mixing matrix appearing in the weak charged current interactions is given by,
\bea
{\mathcal U}^{} &=& \left(
\begin{array}{ccc}
1   & 0 & 0 \\  0 & c_{23}  & s_{23}   \\ 
 0 & -s_{23} & c_{23} \\
\end{array} 
\right)   
  \left(
\begin{array}{ccc}
c_{13}  &  0 &  s_{13} e^{- i \delta}\\ 0 & 1   &  0 \\ 
-s_{13} e^{i \delta} & 0 & c_{13} \\
\end{array} 
\right)  \left(
\begin{array}{ccc}
c_{12}  & s_{12} & 0 \\ 
-s_{12} & c_{12} &  0 \\ 0 &  0 & 1  \\ 
\end{array} 
\right)  \ ,
\label{u}
 \eea 
where $s_{ij}=\sin {\theta_{ij}}, c_{ij}=\cos \theta_{ij}$ and $\delta$ is the Dirac-type CP phase. The form of ${\cal U}$ given in Eq.~\ref{u}  is referred to as the  Pontecorvo-Maki-Nakagawa-Sakata (PMNS) parametrization~\cite{Beringer:1900zz}.
  If neutrinos are Majorana particles, there can be 
two additional Majorana-type phases in the three flavour case. However, those Majorana phases play no role in neutrino oscillation studies. We have measured  the  parameters entering the neutrino oscillation framework to a fairly good precision (see  the global fit analyses~\cite{Capozzi:2017ipn,deSalas:2017kay,globalfit,Esteban:2018azc,nufit}). The best-fit values and $3\sigma$ range of neutrino mass and mixings deciphered from oscillation data are given in Table~\ref{tab:parameters}. 
Yet, there are some unknowns in the standard mass-induced oscillation framework. These
 include the question of neutrino mass hierarchy (sign of $\Delta m^{2}_{31}$), the value of the CP violating phase ($\delta$) and determining the correct octant of
  $\theta_{23}$.  

On the theoretical side, neutrino oscillations require non-zero masses while neutrinos are massless in the SM. This implies that one needs to go beyond the SM in order to explain the results of oscillation experiments. The minimal way is to have a new physics model which can give rise to nonzero neutrino masses but the interactions are still described by SM.
Once we invoke new physics to accommodate neutrino masses, it is only natural to consider the possibility that the neutrino interactions are described by NSI (as was proposed by Wolfenstein~\cite{Wolfenstein:1977ue}). Clearly, a dominant contribution from such interactions is ruled out by the present data~\cite{deSalas:2017kay,globalfit,Esteban:2018azc,nufit}. 
However a subdominant contribution cannot be ruled out given the present accuracy of the neutrino oscillation experiments. Therefore, the idea proposed by Wolfenstein does not hold true in totality in the current times  yet his insight remains in the form of subdominant effects due to NSI on neutrino oscillations. 

The fact that parameter degeneracies crop up in the presence of standard interactions (SI) has been well recognized since the past two decades or so~\cite{Barger:2001yr,Gandhi:2004bj,Huber:2005ep,Hagiwara:2005pe,Kajita:2006bt, Ghosh:2015ena}. Identification and resolution of parameter degeneracies is crucial for a clean determination of the oscillation parameters. 
Besides, any new physics sector (such as NSI considered in the present work) introduces a multitude of parameter degeneracies apart from those in the standard case and the structure of parameter degeneracies is far more complex. There has been a vast body of work done on NSI and neutrino oscillations. For a comprehensive recent review on the topic of NSI in the context of neutrino oscillations, we refer the reader to~\cite{Farzan:2017xzy}. The idea of subdominant NSI in neutrino propagation affecting the CP violation studies, neutrino mass hierarchy and octant of $\theta_{23}$ at upcoming long baseline neutrino experiments has received tremendous attention in neutrino physics in the recent years~\cite{Masud:2015xva, deGouvea:2015ndi, Coloma:2015kiu, Liao:2016hsa, Forero:2016cmb, Huitu:2016bmb, Bakhti:2016prn, Masud:2016bvp, Soumya:2016enw, Rashed:2016rda, Coloma:2016gei, Babu:2016fdt, deGouvea:2016pom, Masud:2016gcl, Blennow:2016etl,Agarwalla:2016fkh, Ge:2016dlx, Forero:2016ghr, Liao:2016bgf, Blennow:2016jkn, Fukasawa:2016lew, Deepthi:2016erc, Liao:2016orc, C.:2017yqh, Rout:2017udo, Ghosh:2017ged, Kelly:2017kch, Shoemaker:2017lzs, Ghosh:2017lim, Farzan:2017xzy, Deepthi:2017gxg,  Wang:2018dwk, Choudhury:2018xsm, Falkowski:2018dmy, Dey:2018yht, Meloni:2018xnk, Flores:2018kwk, Hyde:2018tqt, Verma:2018gwi, Chatterjee:2018dyd, Bischer:2018zcz}  mainly because our ability to search for subdominant effects has increased substantially due to the precisely designed experiments.

 Some of the important long baseline experiments considered are Tokai to Kamioka (T2K)~\cite{Abe:2013hdq}, Tokai to Hyper Kamiokande (T2HK)~\cite{Abe:2015zbg}, Tokai to Hyper Kamiokande with a second detector in Korea (T2HKK)~\cite{Abe:2016ero}, NuMI Off-axis $\nu_e$ Appearance
(No${\nu}$A)~\cite{Ayres:2004js}, Deep Underground Neutrino Experiment (DUNE)~\cite{Acciarri:2015uup, Acciarri:2016ooe}, Long baseline neutrino oscillation (LBNO)~\cite{Agarwalla:2013vyc}.

In order to set the stage for the present work, we summarize the most relevant references dealing with the issue of propagation NSI at long baselines and  constraining NSI parameters.
By carrying out detailed simulations of the DUNE experiment in the presence of new physics, the authors of~\cite{deGouvea:2015ndi} focus on whether DUNE would be able to distinguish between different kinds of new physics such as propagation NSI and sterile neutrino. 
In ~\cite{Coloma:2015kiu}, it was shown that DUNE will improve the constraints over some of the propagation NSI parameters by carrying out sensitivity studies and suggested that a combination of DUNE and T2HK would help in resolving degeneracies among standard and NSI parameters.
Reference \cite{Huitu:2016bmb} focuses on LBNO and addresses prospects of probing strength of propagation NSI parameters at long baseline experiments 
as a function of the oscillation channel, baseline length and detector mass. Correlations between propagation NSI and source as well as detector NSI have been studied in \cite{Blennow:2016etl}.
%MOMENT
Reference \cite{Fukasawa:2016lew} deals with yet another long baseline experiment, T2HKK and discusses how different configurations of T2HKK would be helpful in constraining the propagation NSI.
Reference \cite{Liao:2016orc} discusses the issue of parameter degeneracies in the presence of propagation NSI and the authors perform a comparison of the potential of DUNE, T2HK and T2HKK in probing some of the NSI parameters.
In~\cite{Kelly:2017kch}, the author considers a combination of information from atmospheric neutrinos and long baseline experiment T2HK and its impact on constraining the NSI parameters. It should be noted that the studies carried out so far on constraining NSI terms on DUNE has invariably utilized the standard low energy (LE) flux that peaks around the first oscillation maximum for $\pme$ \ie, around $2-3$ GeV. We advance in this direction by incorporating different beam tunes at DUNE and understand the role of beam tunes in constraining the NSI parameters. 
In a recent work, high energy beams have been shown to be helpful in distinguishing the NSI scenario from the standard three neutrino 
scenario~\cite{Masud:2017bcf}. 
While the new physics context of the present study is that of propagation NSI, our approach is valid for a variety of new physics models. 
%

%-------------------------------------------
\begin{table}[h]
\centering
\begin{tabular}{  c c l }
%\hline
\hline
&&\\
{\textsc{Oscillation Parameter}} & {\textsc{Best-fit value}} &  {\textsc{$3 \sigma$ range}}  \\
&&\\
\hline
&&\\
$\theta_{12}$ [$^\circ$] & $34.5$  &  31.5 - 38.0 \\
$\theta_{13}$ [$^\circ$]  & $8.45$  & 8.0 - 8.9 \\
$\theta_{23}$ [$^\circ$] & $47.7$  & 41.8 - 50.7 \\
$\delta/\pi$ & $ -0.68 $ & $[-1,-0.06]$ ~ {\textrm{and}}~ $[0.87,1]$\\
$\Delta m^2_{21}$ [$10^{-5} \text{ eV}^2$]  & $7.55 $ & 7.05 - 8.14 \\
$\Delta m^2_{31}$ [$10^{-3} \text{ eV}^2$] & $+2.50 $ & 2.41 - 2.60  \\
&&\\
\hline
\end{tabular}
\caption{\label{tab:parameters}
Neutrino mass and mixing parameters obtained from the global fit to neutrino oscillation 
 data~\cite{deSalas:2017kay,globalfit}}.
\end{table}
%-------------------------------------------  

The article is organised as follows.  
In Sec.~\ref{sec:nsi}, we  give the theoretical introduction to neutral current (NC) NSI which is the new physics scenario considered in the present article. We also mention the present constraints on the NSI terms. 
In Sec.~\ref{sec:simulation}, we describe the numerical simulation procedure as well as introduce the beam tunes  used. 
In subsection \ref{sec:prob_basic}, we first discuss the impact of individual NSI terms on the behaviour of probabilities ($\pme$ and $\pmm$) as functions of $\delta$.
In subsections \ref{sec:prob_heatmap_energy} and \ref{sec:prob_heatmap_del}, we 
analyze the behaviour of the probability difference between NSI and SI as a function of energy as well as $\delta$. 
In Sec.~\ref{sec:chisq}, we do a comparative $\Delta \chi^{2}$ analysis to discuss in detail how the higher energy beams in conjunction with the standard low energy beam impact the sensitivities of parameters. 
Finally, we summarize our conclusions in Sec.~\ref{sec:conclude}.
In Appendices~\ref{sec:pme_analysis} and \ref{sec:pmm_analysis}, we have given the relevant probability expressions that aid in understanding our results. Appendix~\ref{sec:event} contains the SI-NSI event
 difference plot for some representative choice of parameters.

%----------------------------------------------------------

\section{Model : Nonstandard interaction during propagation}
\label{sec:nsi}

The new physics scenario considered in the present work is that of  propagation NSI which impacts the propagation of neutrinos. Such a scenario can be described by a dimension-six operator involving four fermions, 
\begin{equation}
{\cal L}_{NSI} = -2 \sqrt 2 G_F \epsilon_{\alpha\beta}^{fC} (\bar \nu_\alpha \gamma^\mu P_L \nu_\beta) (\bar f \gamma_\mu P_C f)
\label{eq:lnsi}
\end{equation}
where $\alpha,\beta = e,\mu,\tau$ indicate the neutrino flavor, $f $ denotes the matter fermions, $e,u,d$.  The new NC interaction terms can impact the neutrino oscillation physics via flavour changing interaction or flavour preserving interaction. 
From a phenomenological point of view, only the sum (incoherent) of all the individual contributions (from different scatterers such as $e$, $u$ or $d$) contributes to the coherent
forward scattering of neutrinos on matter. 
Normalizing to $n_e$, the effective NSI parameter for neutral Earth matter\footnote{For neutral  Earth matter, there are two nucleons (one proton and one neutron) per
  electron. 
  } is given by
\bea
\varepsilon_{\alpha\beta} &=& \sum_{f=e,u,d} \dfrac{n_f}{n_e}
\varepsilon_{\alpha\beta}^f = \varepsilon_{\alpha\beta}^e +2
\varepsilon_{\alpha\beta}^u + \varepsilon_{\alpha\beta}^d + \dfrac{n_n}{n_e}
(2\varepsilon_{\alpha\beta}^d + \varepsilon_{\alpha\beta}^u) = \varepsilon
^e_{\alpha\beta} + 3 \varepsilon^u_{\alpha \beta} + 3
\varepsilon^d_{\alpha\beta} \ ,
      \label{eps_combin}
\eea 
where $n_f$ is the density of fermion $f$ in medium crossed by the
neutrino and $n$ refers to neutrons.  Also, $\varepsilon_{\alpha\beta}^f=
\varepsilon_{\alpha\beta}^{fL} + \varepsilon_{\alpha\beta}^{fR}$ which
encodes the fact that NC type NSI matter effects are sensitive to the
vector sum of NSI couplings.

In the presence of NSI, the  Hamiltonian in the effective \sch-like equation governing neutrino evolution can be expressed as
 \begin{eqnarray}
 \label{hexpand} 
 {\mathcal
H}^{}_{\mathrm{}} &=& 
\dfrac{1}{2 E} \Bigg\{ {\mathcal U} \left(
\begin{array}{ccc}
0   &  &  \\  &  \Delta m^2_{21} &   \\ 
 &  & \Delta m^2_{31} \\
\end{array} 
\right) {\mathcal U}^\dagger 
 + {a (x)}   \left(
\begin{array}{ccc}
1+ \varepsilon_{ee}  & \varepsilon_{e \mu}  & 
\varepsilon_{e \tau}  \\ {\varepsilon_{e\mu} }^ \star & 
\varepsilon_{\mu \mu} &   \varepsilon_{\mu \tau} \\ 
{\varepsilon_{e \tau}}^\star & {\varepsilon_{\mu \tau}}^\star 
& \varepsilon_{\tau \tau}\\
\end{array} 
\right) \Bigg\}  \ ,
 \end{eqnarray} 
where $\Delta m^2_{ij}$ are the mass-squared differences. Here {{$a (x)= 2 \sqrt{2}  	E G_F n_e (x)$}} is the standard charged current (CC) potential due to
the coherent forward scattering of neutrinos, $n_e$ is the electron
number density and ${\varepsilon}_{\alpha \beta} \, (\equiv |\varepsilon _{\alpha \beta}|
\, e^{i \varphi_{\alpha\beta}})$ are complex NSI parameters.
${\cal U}$ is the PMNS three flavour neutrino mixing matrix  (see Eq.~\ref{u}).

We now mention the constraints on the NC NSI parameters.  The combination that
enters oscillation physics is given by
Eq.~\ref{eps_combin}.  Assuming that the
errors on individual NSI terms are uncorrelated, 
model-independent bounds on NC NSI terms $\varepsilon_{\alpha \beta}$ were given in Ref.~\cite{Biggio:2009nt}. In particular, one obtains the following:
\begin{eqnarray}
\varepsilon_{\alpha\beta} \lsim \left\{  \sum_{C=L,R} [ (\varepsilon_{\alpha \beta}^{e C} )^2 + (3 \varepsilon_{\alpha\beta}^{u C})^2 + (3 \varepsilon _{\alpha \beta}^{d C})^2 ] \right\}^{1/2} \ ,
\end{eqnarray} 
 which leads to 
 \begin{eqnarray}
 |\varepsilon_{\alpha\beta}|
 \;<\;
  \left( \begin{array}{ccc}
4.2  &
0.33 & 
3.0 \\
0.33 &0.068 & 0.33 \\
3.0  &
0.33 &
21 \\
  \end{array} \right) \ .\label{largensi}
\end{eqnarray} 
 for neutral Earth matter.
Direct experimental constrains from neutrino experiments on NSI parameters are more restrictive. 
The SK NSI search in atmospheric neutrinos crossing the Earth found no evidence
in favour of NSI and the study led to upper bounds on NSI
parameters~\cite{Mitsuka:2011ty} given by $|\varepsilon_{\mu\tau}| < 0.033, |
\varepsilon_{\tau\tau} - \varepsilon_{\mu\mu} | < 0.147 $ (at 90\% CL) in a
two flavour hybrid model~\cite{Ohlsson:2012kf}.  The off-diagonal NSI parameter
$\varepsilon_{\mu\tau}$ is constrained $-0.20 < \varepsilon_{\mu\tau} <
0.07$ (at 90\% CL) from MINOS data in the framework of two flavour
neutrino oscillations~\cite{Adamson:2013ovz,Kopp:2010qt}.

In what follows, we shall adopt a numerical approach to discuss the impact of various NSI parameters. For the sake of simplicity and clarity, we consider one NSI parameter at a time. Wherever analytic description is feasible, we give approximate analytic expressions which are valid in the present context and additional plots which help in understanding the results obtained numerically (for more details, see Appendices~\ref{sec:pme_analysis} and \ref{sec:pmm_analysis}).

%----------------------------------------------------------

\section{Simulation procedure  and beam tunes}
\label{sec:simulation}
The proposed Deep Underground Neutrino Experiment (DUNE)  has a baseline of 1300 km and a 40 kt liquid argon far detector (FD) is placed at an on-axis location. 
The primary scientific goals of 
DUNE include the measurement of leptonic CP violation, the determination of the neutrino mass ordering and the precision 
measurement of the neutrino mixing parameters~\cite{Acciarri:2015uup, Acciarri:2016ooe, Acciarri:2016crz, Abi:2018dnh}. 

In order to simulate DUNE, we use the GLoBES package \cite{Huber:2004ka, Huber:2007ji} with the most recent DUNE configuration file provided by 
the collaboration \cite{Alion:2016uaj} and implement the density profile given by Preliminary Reference Earth Model (PREM)~\cite{Dziewonski:1981xy}. 
We assume a total runtime of 7 years with  3.5 years  in the neutrino mode and another 3.5 years in the antineutrino mode. 
\begin{table}[htb!]
\centering
{{
\begin{tabular}{| l | l l l|}
\hline
%&&&\\
Parameter & LE & ME & \\
%&&&\\
\hline \hline
%&&&\\
Proton beam  & $E_{p^+} =80$ GeV &$ E_{p^+} =120$ GeV & \\
  & 1.07 MW & 1.2 MW & \\
Focusing &  \multicolumn{3}{c|}{2 NuMI horns, 230kA, {{6.6 m apart}} }\\
Target location & {{-25 cm}} & {{-1.0 m}} &  \\
Decay pipe length & 204 m & 250 m  &  \\
Decay pipe diameter & 4 m & 4 m &   \\
\hline
\end{tabular}
}}
\caption{\label{tab:cdr} 
Beamline parameters assumed for the different design fluxes used  in our sensitivity calculations~\cite{2013arXiv1307.7335L, Alion:2016uaj}. The target is a thin Be cylinder 2 interaction lengths long. The target location is given with respect to the upstream face of Horn 1. {The LBNF neutrino beamline decay pipe length has been chosen to be 194 m. Decay pipe lengths of up to 250 m could be accommodated on the Fermilab site and were an option in previous designs of the beamline.}
} 
\end{table}
\begin{figure}[htb]
\centering
\includegraphics[width=.5\textwidth,height=.45\textwidth]
{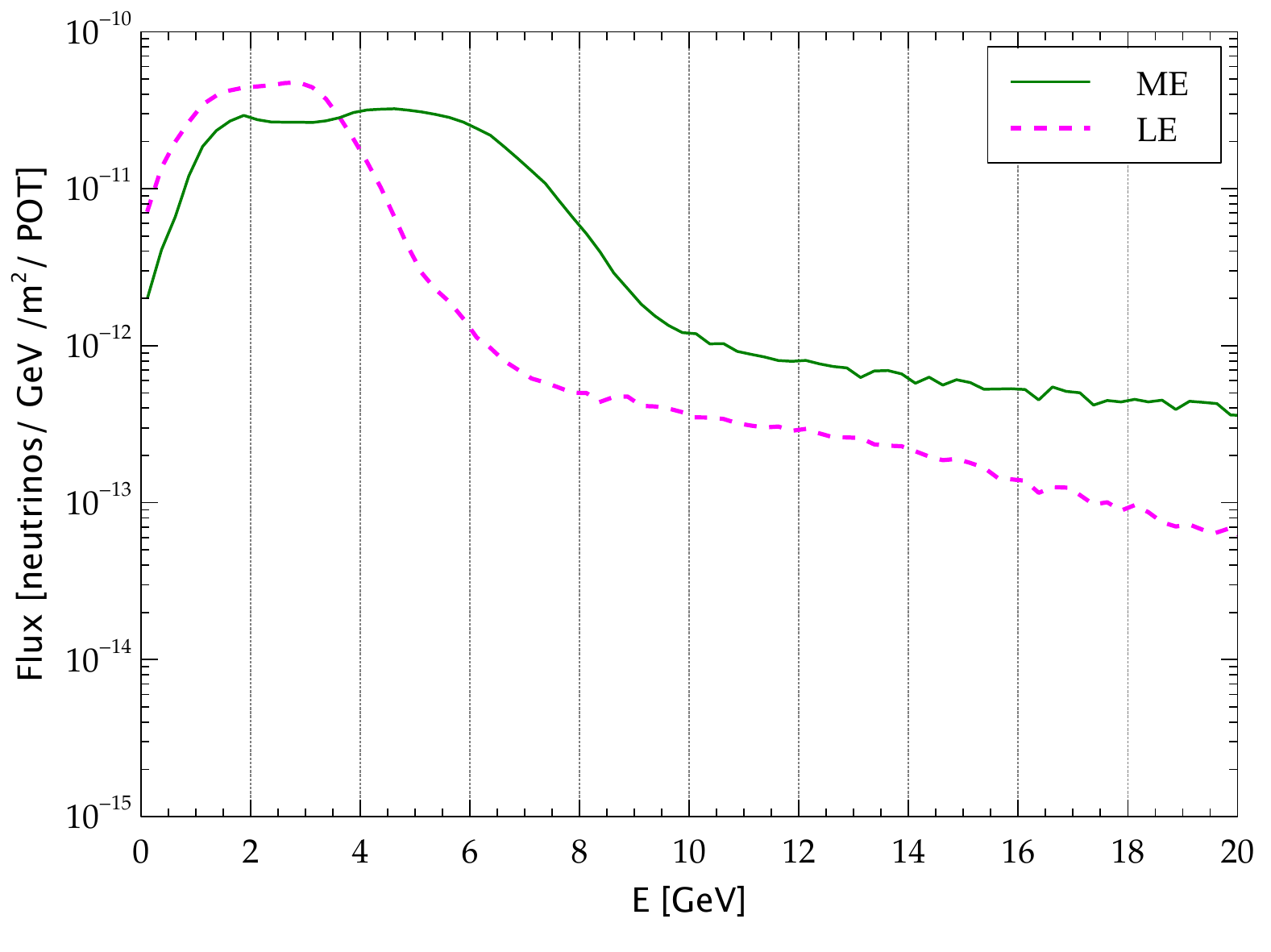}
\caption{\footnotesize{The  neutrino fluxes (LE and ME) used in the present work.
LE beam refers to the standard flux generated by an 80 GeV proton beam as used in 
\cite{Alion:2016uaj}. ME beam refers to the flux peaking at a higher energy.
See Table~\ref{tab:cdr} for more details. 
}}
\label{fig:flux_comp}
\end{figure}

We consider two 
beam tunes obtained from a G4LBNF simulation~\cite{Agostinelli:2002hh, Allison:2006ve} of the LBNF beam line using NuMI-style focusing. 
\begin{description}\item[LE beam] 
The standard $\nu_{\mu}$ beam which peaks around a relatively lower energy  of $\sim 2.5$ GeV (corresponding to the first oscillation maximum for the $\nu_{\mu} \to \nu_{e}$ appearance channel) is referred to as an LE beam in our analyses. 
It is generated by an $80$ GeV proton beam delivered at 1.07 MW with protons on target (pot) of $1.47 \times 10^{21}$.
\item[ME beam] 
The second beam is has the characteristic that it is larger at higher energies ($\gtrsim 4$ GeV onwards) and we refer to this beam as medium energy (ME) beam. 
The ME beam is generated by a 120 GeV proton beam delivered at 1.2 MW with a pot  of $1.1 \times 10^{21}$. 
\end{description}
Both the LE and ME fluxes are shown in Fig.~\ref{fig:flux_comp}.  
The LE flux peaks around 1.5 GeV to 3.5 GeV but after that it falls off  rapidly. In contrast, the ME flux is almost flat from 2 - 6 GeV and after that it falls off but at a much slower rate compared to the LE flux and it remains substantially higher than the LE flux even beyond 6 GeV. 
At $\sim 2.5$ GeV, the ME flux is $\sim 25-35\%$ smaller than the LE flux. 
Hence, in our analyses of probing the NSI parameters, we use a combination of LE and ME flux together, so as to extract information on new physics from both the lower energy ($1-3$ GeV) and the higher energy ($\gtrsim 4$ GeV) regime as much as possible.
We compare the results with those obtained using the LE beam only for the same total runtime of the experiment. 
The beamline parameters assumed for the different design fluxes used in our sensitivity
calculations are given in Table~\ref{tab:cdr}.

Our analysis includes both appearance ($\nu_{\mu} \to \nu_{e}$)  and  disappearance ($\nu_{\mu} \to \nu_{\mu}$)
channels, simulating both signal as well as background. 
The simulated background includes contamination of antineutrinos (neutrinos) in the neutrino (antineutrino) mode, and also misinterpretation of flavors, 
as discussed in detail in \cite{Alion:2016uaj}.
To analyze the NSI scenario, we utilise the GLoBES extension called 
{\textbf{snu.c}} which is described in~\cite{Kopp:2006wp, Kopp:2007ne}.

To calculate the sensitivity with which the NSI parameters can be probed, one can define     
 the (statistical) $\Delta \chi^2$ as follows~\footnote{The definition of the $\Delta \chi^{2}$ 
 in Eq.~\ref{eq:chisq_si_nsi} includes only statistical effects for the purpose of understanding. 
The systematic effects have of course been taken into account in our numerical results obtained using GLoBES.}:
 \begin{equation}
 \label
 {eq:chisq_si_nsi}
\Delta \chi^2 \simeq \sum_{i}^{\textrm{}} \sum_{j}^{\textrm{}} \dfrac{[N^{ij}_{\textrm{true}}({\rm{SI}}) - N^{ij}_{\textrm{test}}({\textrm{NSI}})]^2}{N^{ij}_{\textrm{true}}({\textrm{SI}})}.
\end{equation}
Here, the SI case is treated as \textit{true} while the NSI parameters are allowed to vary in the \textit{test} dataset.  
The sum over the number of channels runs over the $\nu_\mu \to \nu_e$ and $\nu_\mu \to \nu_\mu$  channels and the corresponding antineutrino channels, $\bar \nu_\mu \to \bar \nu_e$ 
and $\bar \nu_\mu \to \bar \nu_\mu$. 
The index $j$ indicates the sum over all the energy bins ranging from $E=0-20$ GeV. We have a total of $71$ bins of non-uniform widths ($64$ bins with uniform bin width of $125$ MeV in the energy range $E =0-8$ GeV and $7$ bins with variable width beyond $8$ GeV) \cite{Alion:2016uaj}.
The detector configuration, efficiencies, resolutions and systematic uncertainties for \dune~ are listed in Table.~\ref{tab:sys}. 

%-------------------------------------------
\begin{table}[ht]
\centering
\begin{tabular}{ |l| l l |ll| }
\hline
Detector details & \multicolumn{2}{c |}{Normalisation error} & \multicolumn{2}{c |}{Energy calibration error} \\
                                         \cline{2-5}
                                        & Signal & Background       & Signal & Background                              
                                        \\ 
\hline
\dune 
&&&&\\
 Runtime (yr) = 3.5 $\nu$ + 3.5 $\bar \nu$ 
  & $\nu_e : 5\%$  & $\nu_e : 10\%$ &   $\nu_e : 2\%$ & $\nu_e : 10\%$\\
40 kton, LArTPC & & &&\\
& $\nu_\mu : 5\%$ & $\nu_\mu : 10\%$ & $\nu_\mu : 5\%$ & $\nu_\mu : 10\%$\\
&&&&\\
\hline
\end{tabular}
\caption{\label{tab:sys} 
Detector configuration, efficiencies, resolutions and systematic uncertainties for \dune.  }
\end{table}
%-------------------------------------------

We have used the standard oscillation parameters in Table~\ref{tab:parameters}, taken from Ref.~\cite{deSalas:2017kay, globalfit}.  
For the neutrino mass hierarchy, we 
assume a spectrum corresponding to normal hierarchy in the true dataset. 
Since DUNE has no sensitivity to the solar parameters and since  
 $\theta_{13}$ is rather well measured by current reactor and long baseline experiments, we keep these values fixed to their current best-fit values, while marginalizing 
over  $\theta_{23}$ (in the present $3\sigma$ range) and $\delta$ ($[-\pi,\pi]$), if not plotting them. 
In addition, we marginalize over the atmospheric mass-squared splitting, $\Delta m_{31}^2$, allowing for the two possible mass hierarchies. 
When studying a non-diagonal NSI parameter, $\varepsilon_{\alpha\beta}$, we also marginalize 
over its corresponding phase, $\varphi_{\alpha\beta}$ in the range $[-\pi,\pi]$. 
Therefore, if we study two non-diagonal complex parameters simultaneously, we marginalize over a total of five parameters.

In our analysis, we consider two diagonal NSI parameters and three off-diagonal NSI parameters with both their moduli and phases. 
If we also include the yet unknown CP phase, $\delta$, we  have a total of nine parameters.  We depict $\Delta \chi^{2}$ correlations among these nine parameters 
($\delta, \varepsilon_{ee}, |\varepsilon_{e\mu}|,\varphi_{e\mu},|\varepsilon_{e\tau}|,\varphi_{e\tau}, 
 |\varepsilon_{\mu\tau}|,  \varphi_{\mu\tau},  \varepsilon_{\tau\tau}$) considering them pairwise at a time and the number of such combinations is 36.

%----------------------------------------------------------

\section{A scan of parameter space at the level of probability}
\label{sec:prob}
 
In order to obtain insight into the correlations and degeneracies among the various NSI and SI parameters that may impact the signals at DUNE, the first step is, naturally, to look at the relevant oscillation probabilities. We consider the following oscillation channels that are accessible\footnote{$\nu_\mu \to \nu_\tau$ is also in principle there, but the signal is extremely tiny.} at DUNE : 
 \begin{enumerate}
\item Appearance channel : $\nu_{\mu} \to \nu_{e}$ 
\item Disappearance channel : $\nu_{\mu} \to \nu_{\mu}$ 
\end{enumerate} 
  \begin{figure}[htb]
\centering
\includegraphics[width=1\textwidth,height=.5\textwidth]
{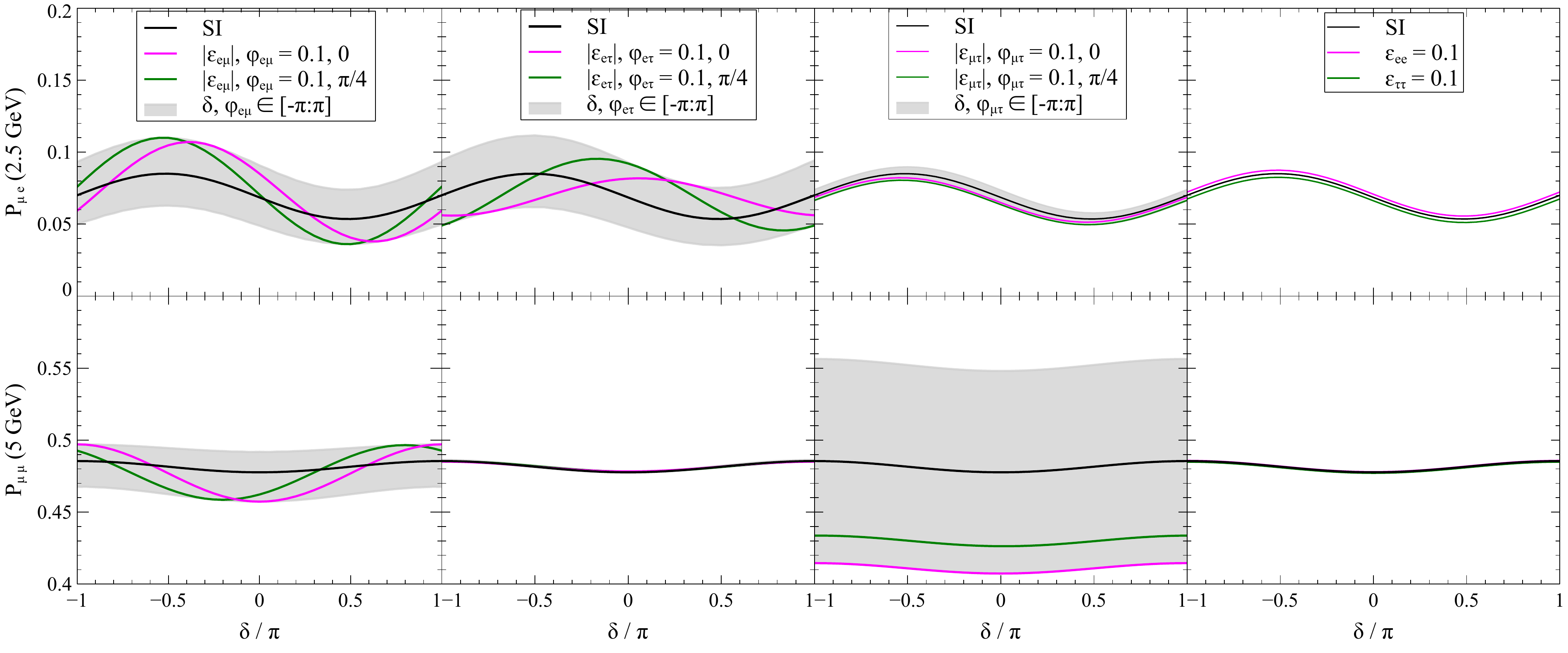}
\caption{\footnotesize{ $\pme$ (top row) and $\pmm$ (bottom row) at fixed baseline ($L=1300$ km) and fixed  energy values ($E=2.5$ GeV for $P_{\mu e}$ and $E=5$ GeV for $P_{\mu\mu}$) plotted as a function of the CP phase, $\delta$.  
The strength of all NSI terms is taken to be the same ($=0.1$).
}}
\label{fig:p_vs_dcp}
\end{figure}

In what follows, we consider the relevant parameters that include two of the diagonal NSI parameters ($\varepsilon_{ee},\varepsilon_{\tau\tau}$) and the moduli and phases of the three off-diagonal NSI parameters ($\varepsilon_{e\mu}$,$\varepsilon_{e\tau}$,$\varepsilon_{\mu\tau}$). 
A detailed assessment of the role of individual NSI terms on the different oscillation channels has been carried out in \cite{Kikuchi:2008vq, Asano:2011nj}. Based on the analyses in \cite{Kikuchi:2008vq, Asano:2011nj}, we can conclude that among all NSI parameters, $\eem$ and $\eet$ mainly impact 
the appearance channel ($\nu_\mu \to \nu_e$)  while $\eee$ has a milder impact. It is clear that $\eem$ enters $\nu_\mu\to\nu_e$ channel.
The almost maximal mixing in the  $2-3$ sector ensures that $\eet$ also impacts this channel with similar strength as $\eem$ (see Appendix~\ref{sec:pme_analysis} and the discussion in Sec. IV of \cite{Kikuchi:2008vq}).  Similarly, the disappearance channel ($\nu_\mu \to \nu_\mu$) is more sensitive to the presence of NSI parameter $\emt$ (see Appendix~\ref{sec:pmm_analysis} and the discussion in Sec. IV of \cite{Kikuchi:2008vq}).

In the following  sub-sections, we perform a scan of the parameter space at the probability level. We first discuss the fixed energy and fixed baseline snapshots of probabilities (subection \ref{sec:prob_basic}). We then discuss SI-NSI degeneracies in the context of DUNE as a function of energy keeping $\delta$ fixed at the best-fit value (subsection \ref{sec:prob_heatmap_energy}).  Further, we go on to the discussion of SI-NSI degeneracies as a function of  $\delta$ (keeping the energy fixed) in subection \ref{sec:prob_heatmap_del}.

%----------------------------------------------------------

\subsection{Snapshots of $\pme$ and $\pmm$ at fixed energy and fixed baseline}
\label{sec:prob_basic}

 In Fig.~\ref{fig:p_vs_dcp}, we fix the baseline at $1300$ km and show the impact of NSI parameters~\footnote{The moduli of all the NSI parameters have been chosen to be equal to $0.1$ (allowed by present constraints~\cite{Biggio:2009nt,Esteban:2018ppq}). For reasons of clarity and  simplicity, we take one NSI parameter non-zero at a time. 
} on snapshots of $\pme$ and $\pmm$ as a function of $\delta$  at certain (appropriately chosen) fixed energy values. 
This aids in identification of parameters that may have the largest impact at the level of probabilities, though at specific energy values. 
For the $\nu_{\mu} \to \nu_e$ channel (top row in Fig.~\ref{fig:p_vs_dcp}), we choose the fixed value of energy to be $E=2.5$ GeV.  
This value corresponds to the first  oscillation maximum for $P_{\mu e}$.
 On the other hand, $\pmm$ is very close to zero at $2.5$ GeV while it is substantial at higher values of energy. 
 Hence we depict curves for $\pmm$ (bottom row in Fig.~\ref{fig:p_vs_dcp}) at $5$ GeV.  
 The grey bands show the variation of the probability when the relevant phases ($\delta, \varphi_{\alpha\beta}$) are allowed to vary in the range $[-\pi, \pi].$
 As a reference, the SI case is shown as a solid black line in all the plots.
 
As far as $\pme$ at $2.5$ GeV (top row) is concerned, we note that the effect of $\eem$ or $\eet$ is more pronounced when compared to the other NSI terms.  
The presence of $\eem$ or $\eet$ modifies the overall amplitude and the location of the peaks/dips of the probabilities  while the presence of a nonzero $\eemp$ or $\eetp$ brings in additional phase shifts. We note that $\emt$ has a much smaller effect on $\pme$. 
$\eee$ and $\ett$ also have a miniscule effect on the amplitude of $\pme$. On the other hand, $\pmm$ at 5 GeV (bottom row) gets affected most by the presence of the $\emt$ term. $\eet, \eee, \ett$ have practically no impact on $\pmm$. 
$\eem$ induces some phase dependence on $\pmm$. 
    
    In what follows, we generalise the above discussion and study the energy dependence of the SI-NSI degeneracies for  $\pme$ and $\pmm$ and also vary the NSI terms instead of keeping their values fixed.

%----------------------------------------------------------

\subsection{Energy dependence of the SI-NSI degeneracies 
}
\label{sec:prob_heatmap_energy}    
        \begin{figure}[htb]
\centering
\includegraphics[width=.9\textwidth,height=.5\textwidth]{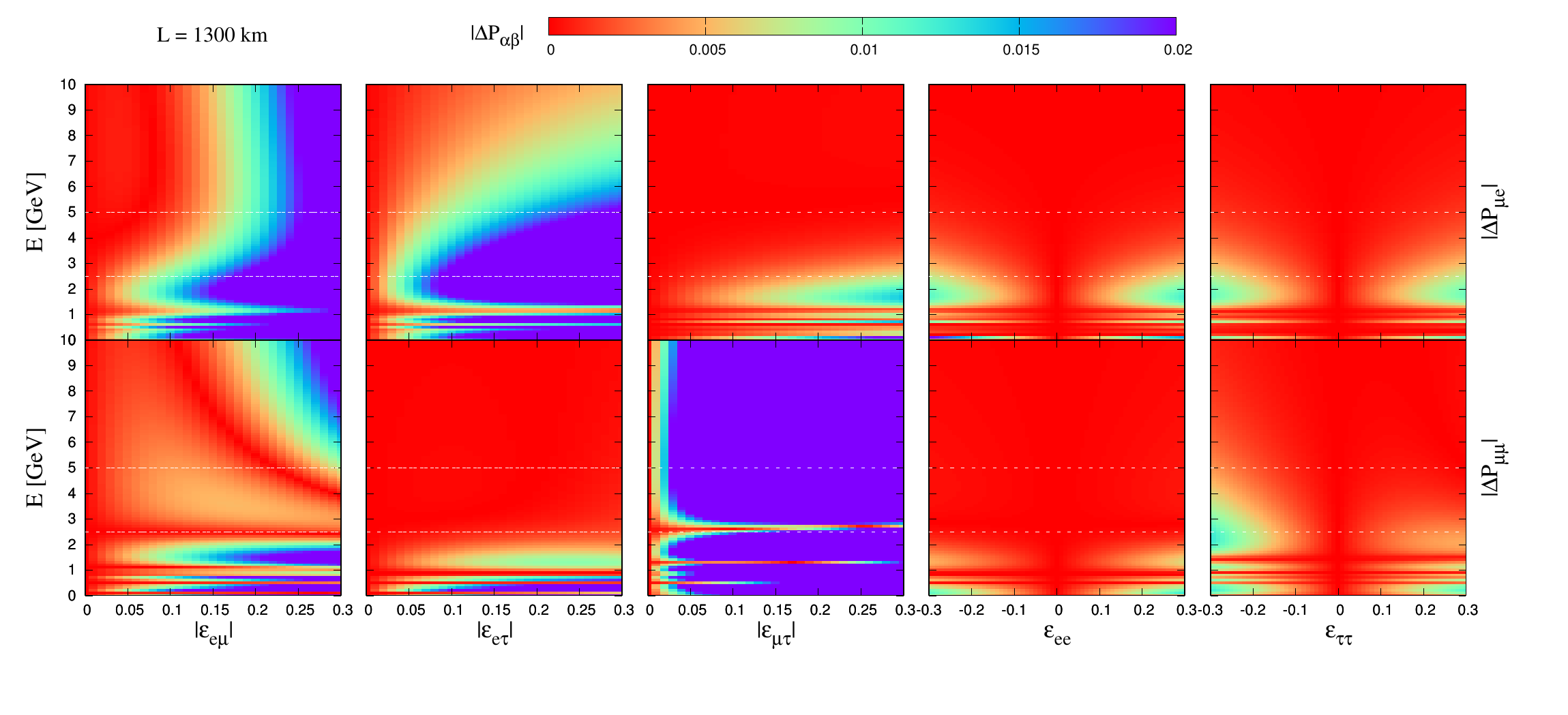}
\caption{\footnotesize{Heatmaps corresponding to $|\Delta \pme|$ (top row) and $|\Delta \pmm|$ (bottom row) 
 in the two-dimensional plane of individual NSI term ($\varepsilon_{\alpha\beta}$) and energy. The NSI phases are set to zero.
The dashed white lines indicate the value of energy at $2.5$ GeV and $5$ GeV.
}}
\label{fig:eps_deg_en}
\end{figure}

To quantify the impact of NSI terms, let us define a quantity, $|\Delta P_{\alpha \beta}| = | P_{\alpha \beta}^{NSI} - P_{\alpha \beta}^{SI}| (\alpha,\beta=e,\mu)$, which is absolute value of probability difference between the SI and NSI scenarios.

\begin{figure}[h!]
\centering
\includegraphics[width=.9\textwidth,height=.5\textwidth]
{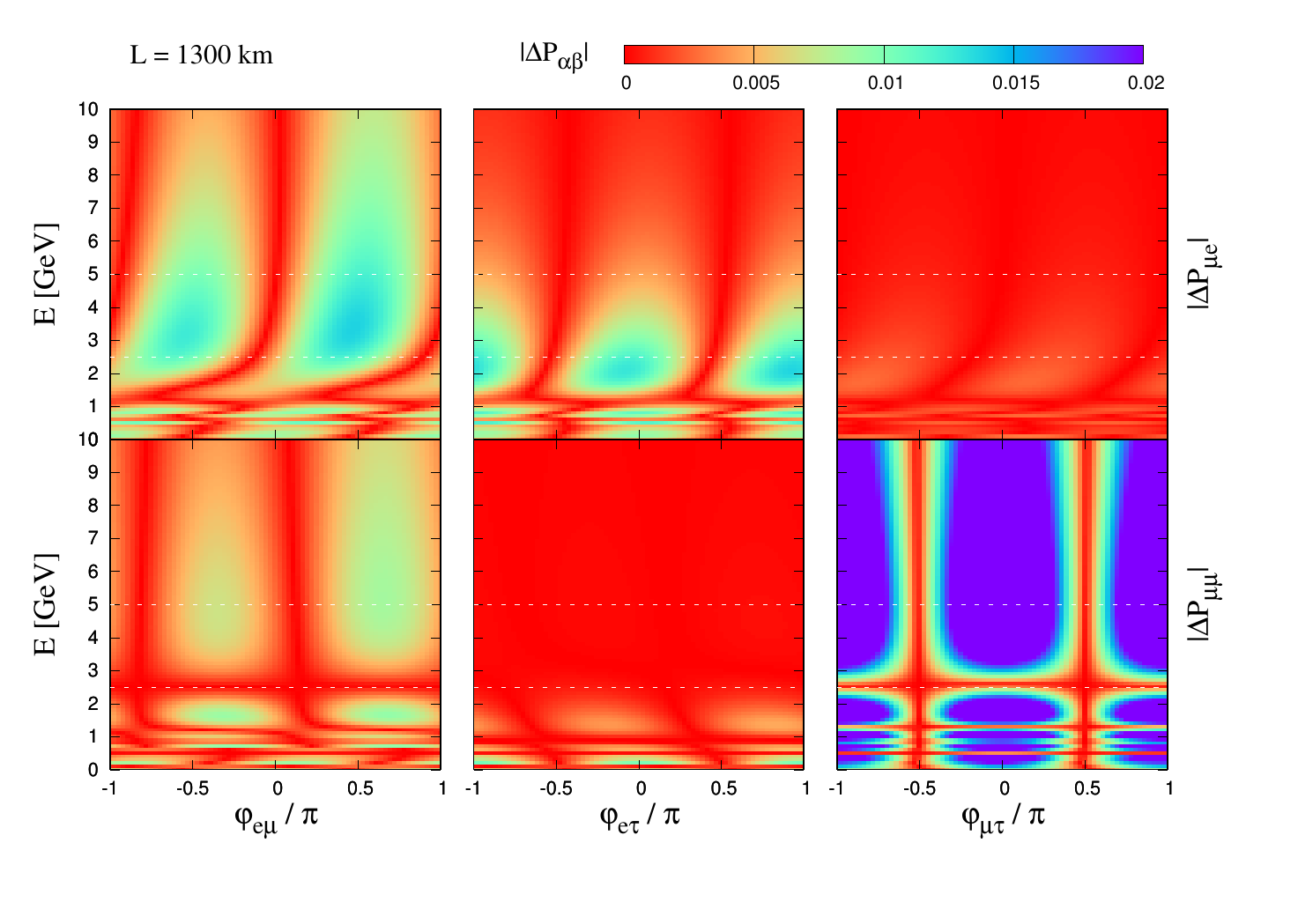}
\caption{\footnotesize{Heatmaps for $|\Delta \pme|$ (top row) and $|\Delta \pmm|$ (bottom row) 
are shown as the function of a NSI phase $\varphi_{\alpha\beta}$ (taken one at a time) 
as the energy is changed, keeping the baseline fixed at 1300 km. 
The associated NSI amplitudes ($\varepsilon_{\alpha\beta}$) were kept fixed at 0.05. The two horizontal  dashed white lines correspond to the energies 2.5 GeV and 5 GeVs.}}
\label{fig:phi_deg_en}
\end{figure}

Our results are given  in  Fig.~\ref{fig:eps_deg_en}  in the form of  heatmaps as functions of energy and the strength of the NSI parameter for $|\Delta \pme|$ (top row) and $|\Delta \pmm|$ (bottom row). The NSI phases are taken to be zero and the standard oscillation parameters have been pinned to their best-fit values (see Table~\ref{tab:parameters}).
If we carefully examine the top row of Fig.~\ref{fig:eps_deg_en}, we note that $|\Delta \pme|$ is mostly affected by $\eema$ and $\eeta$.  
Note that, the impact of $\eema$ or $\eeta$ is most prominent around $2-3$ GeV. 
One can derive a useful conclusion here regarding difference in impact of $\eema$ and $\eeta$ on $|\Delta \pme|$. As we go beyond $\sim 4$ GeV, $\eeta$ gradually makes $|\Delta \pme|$ smaller (red region), while $\eema$ makes $|\Delta \pme|$ stay at a high value (blue region) which is almost independent of energy.
This, in turn, suggests that one may be able to probe $\eem$ more effectively than $\eet$ by use of higher energy beam tune. The other NSI terms $\emt, \eee$ or $\ett$ do not induce much change, keeping $|\Delta \pme| \lesssim 0.005$ for most 
of the energy range.

From the bottom row of Fig.~\ref{fig:eps_deg_en} corresponding to $|\Delta \pmm|$, 
we note that $\emt$ plays an important role. $|\Delta \pmm|$ is large (blue) in most of the energy range as long as  
 $|\emt|$ $\gtrsim 0.02$. This is to be contrasted with other NSI terms, as even a small value of $\emt$ can induce a large impact on $|\Delta P_{\mu\mu}|$. 
 As can be noted, a higher energy beam may be able to probe $\emt$ via this channel effectively.
 If we look at the impact of $\eema$ and $\eeta$, we note that  $\eema$ gradually makes $|\Delta \pmm|$ larger at $E \gtrsim 5$ GeV (indicated by blue region on the  top right side of the panel) while $\eeta$ does not seem to impact $|\Delta \pmm|$. Thus, for the disappearance channel as well, it appears that the higher energy beam may prove more useful in probing $\eem$ than $\eet$. For an analytic understanding of the energy dependence of $\Delta P_{\alpha \beta}$ in the presence of NSI, see Appendices~\ref{sec:pme_analysis} and \ref{sec:pmm_analysis}.

We next consider the case of nonzero phases $\varphi_{\alpha\beta}$. In Fig.\ \ref{fig:phi_deg_en}, heatmaps corresponding to $|\Delta \pme|$ (top row) and $|\Delta \pmm|$ (bottom row) 
 in the two-dimensional plane of individual NSI phase ($\varphi_{\alpha\beta}$) and energy are shown. The moduli of NSI terms ($\eema, \eeta$ or $\emta$) were kept fixed at $0.05$. From Fig.\ \ref{fig:phi_deg_en}, we note that $|\Delta \pme|$ (top row) is most affected by $\eemp$ or $\eetp$ while $\emtp$   has almost no effect. 
Around $ 2-4$ GeV, $\eemp$ and $\eetp$ produce similar qualitative features indicating SI-NSI degeneracy (red band)  occurring at a pair of values given by $\eemp \approx 0,\pm \pi$ and $\eetp \approx -0.6\pi, 0.4\pi$. 
At energies beyond 4 GeV, $|\Delta \pme|$ is very small ($\sim 0$) almost uniformly in the presence of $\eetp$ 
while in the presence of $\eemp$, it still exhibits a relatively high value ($0.005 - 0.01$) in large region of the parameter space. 
 For a quantitative understanding of this feature, we refer the reader to Appendix~\ref{sec:pme_analysis}.

For $|\Delta \pmm|$, Fig.\ \ref{fig:phi_deg_en} (bottom row) shows that it is affected most significantly by $\emtp$,  showing sharp SI-NSI degeneracy around $\emtp \approx \pm \pi/2$.  
This arises because of the fact that $|\Delta \pmm| \propto \cos \emtp$  
(See Eq.~\ref{eq:pmm_emt} in Appendix~\ref{sec:pmm_analysis}). 
We also note that
  $|\Delta \pmm|$ remains close to zero in the presence of $\eetp$ and shows moderate variation for in the presence of $\eemp$.

Finally, we would like to mention that the qualitative features of Fig.\ \ref{fig:phi_deg_en} remain unchanged even if the moduli of the relevant off-diagonal NSI terms ($\eema, \eeta$ or $\emta$) are increased.

%----------------------------------------------------------

\subsection{$\delta$-dependence of SI-NSI degeneracies}
\label{sec:prob_heatmap_del}
   
In Figs.~\ref{fig:pme_eps_deg_del} and \ref{fig:pmm_eps_deg_del} we depict the heatmaps for $|\Delta \pme|$  and  $|\Delta \pmm|$ in the $\delta-\varepsilon_{\alpha\beta}$ plane. In these plots, the first (second) row corresponds to a fixed energy of $2.5$ ($5$) GeV. We can derive the following  conclusions in connection with $\pme$ (see also Appendix~\ref{sec:pme_analysis}) :
\begin{figure}[htb]
\centering
\includegraphics[width=.9\textwidth,height=.5\textwidth]{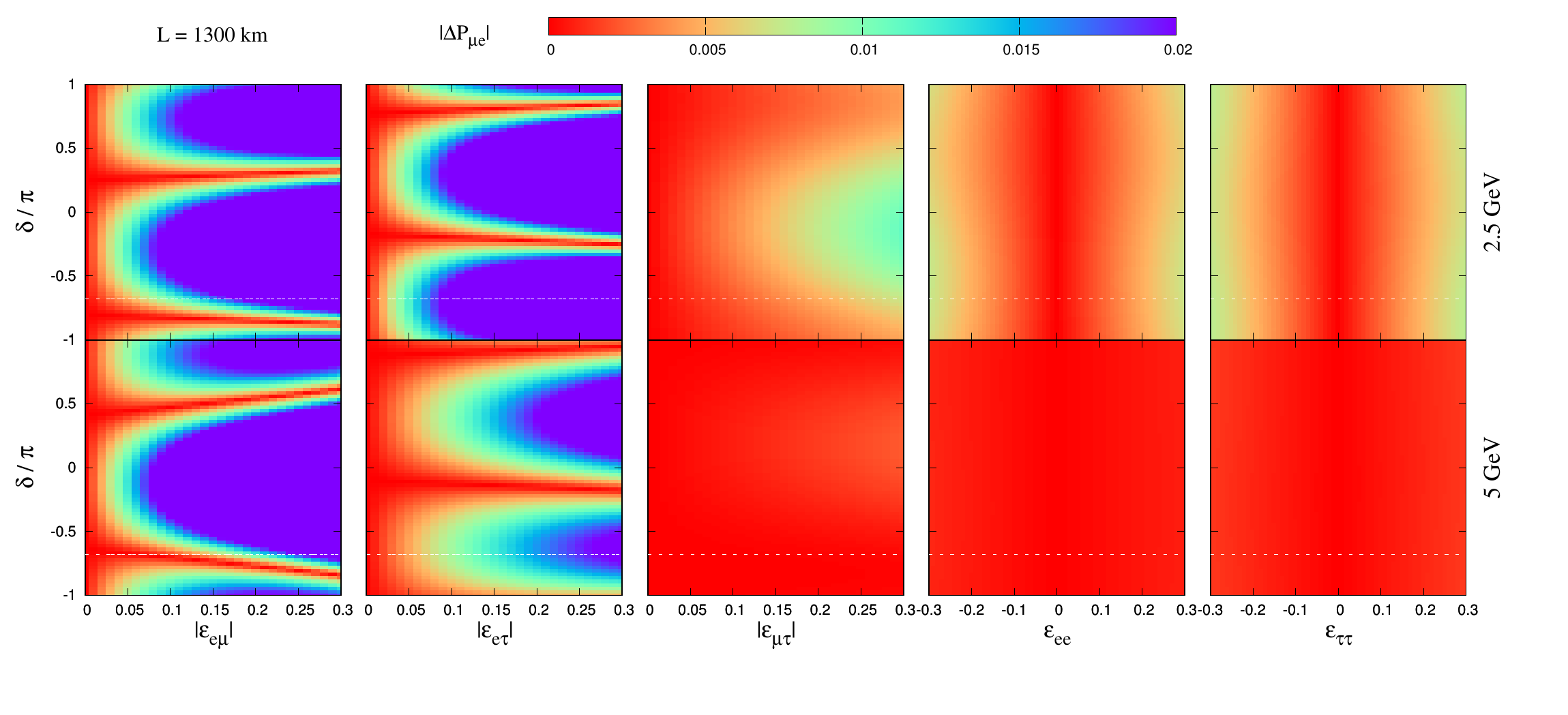}
\caption{\footnotesize{Heatmaps for $|\Delta \pme|$ 
are shown for a fixed baseline of 1300 km in the parameter space of $\delta-\varepsilon_{\alpha\beta}$ for two fixed energies: 2.5 GeV (top row) and 5 GeV (bottom). A single NSI parameter was considered at a time and the associated NSI phases were taken to be zero. 
The dashed horizontal white line corresponds to the bestfit value of the Dirac CP phase $\delta$ ($\approx -0.68\pi$) taken from Table~\ref{tab:parameters}.}}
\label{fig:pme_eps_deg_del}
\end{figure}

\begin{figure}[htb]
\centering
\includegraphics[width=.9\textwidth,height=.5\textwidth]{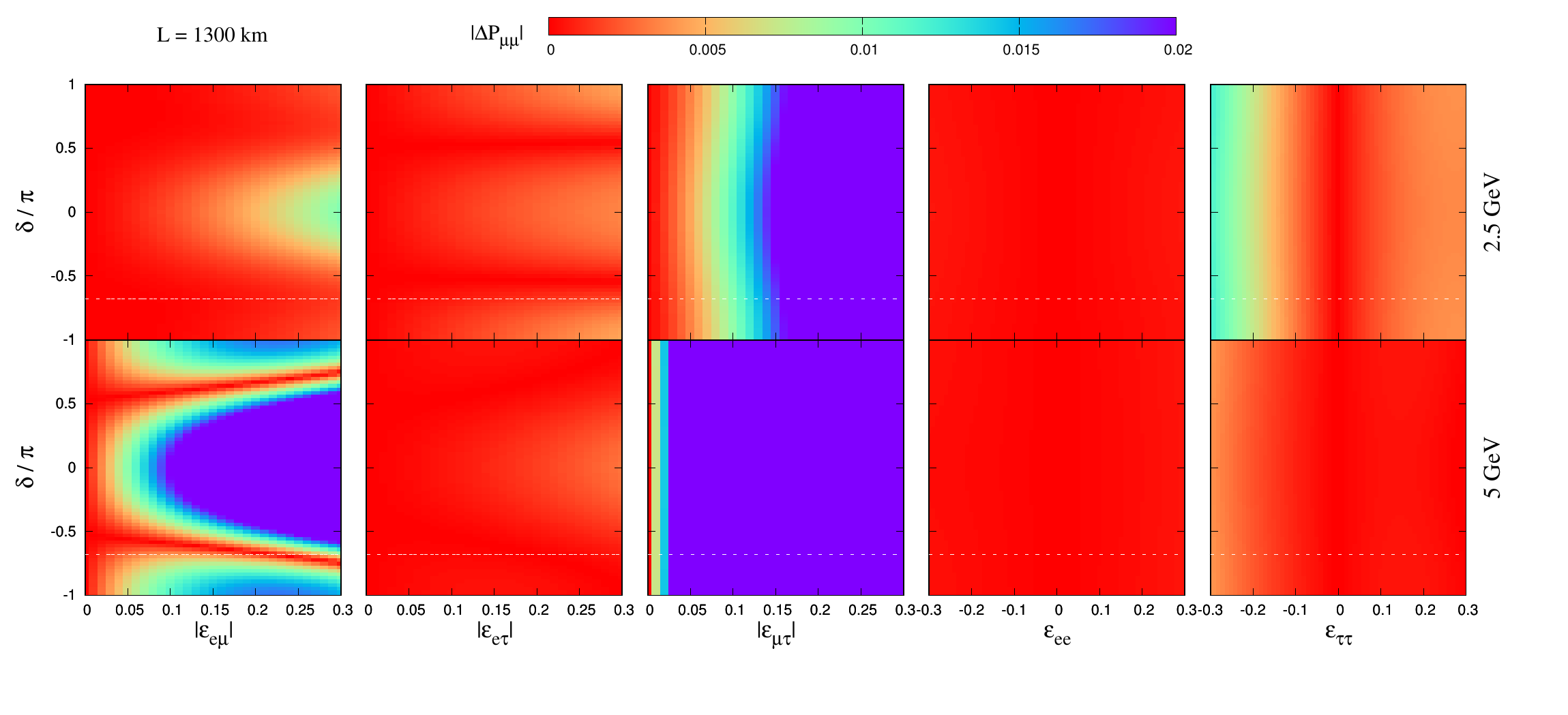}
\caption{\footnotesize{Similar to Fig.~\ref{fig:pme_eps_deg_del} but for the $\nu_{\mu} \rightarrow \nu_{\mu}$ channel.}}
\label{fig:pmm_eps_deg_del}
\end{figure}

\begin{itemize}

\begin{figure}[h!]
\centering
\includegraphics[width=.9\textwidth,height=.5\textwidth]{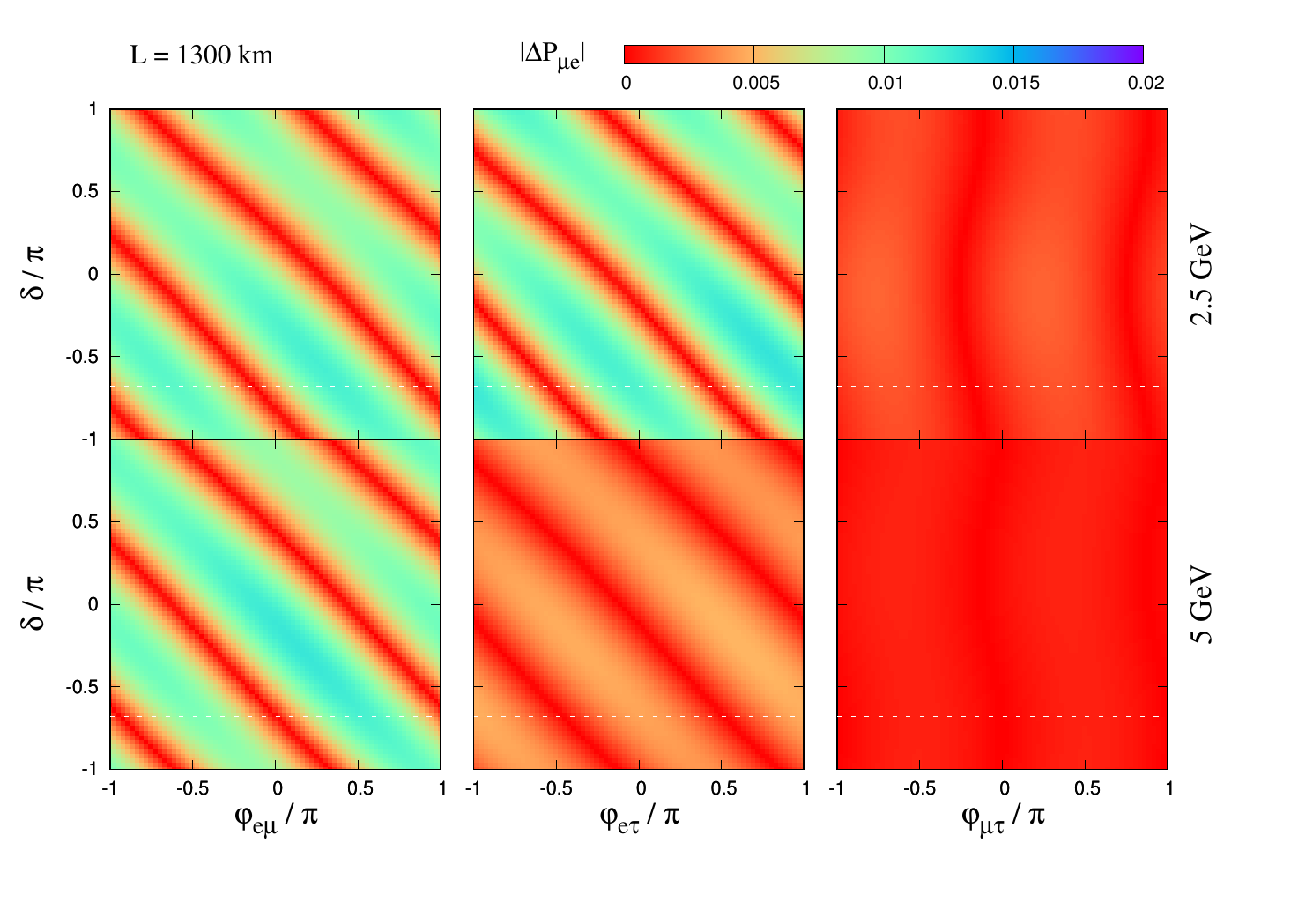}
\caption{\footnotesize{Heatmaps for $|\Delta \pme|$ 
are shown for a fixed baseline of 1300 km in the parameter space of $\delta-\varphi_{\alpha\beta}$ for two values of energies, $2.5$ GeV (top row) and $5$ GeV (bottom row).  Note that $|\varepsilon_{\alpha\beta}|$ was fixed to $0.05$. 
The dashed horizontal white line corresponds to the bestfit value of the Dirac CP phase $\delta$ ($\approx -0.68\pi$) taken from Table~\ref{tab:parameters}.
}}
\label{fig:pme_phi_deg_del}
\end{figure}

\begin{figure}[htb]
\centering
\includegraphics[width=.9\textwidth,height=.5\textwidth]{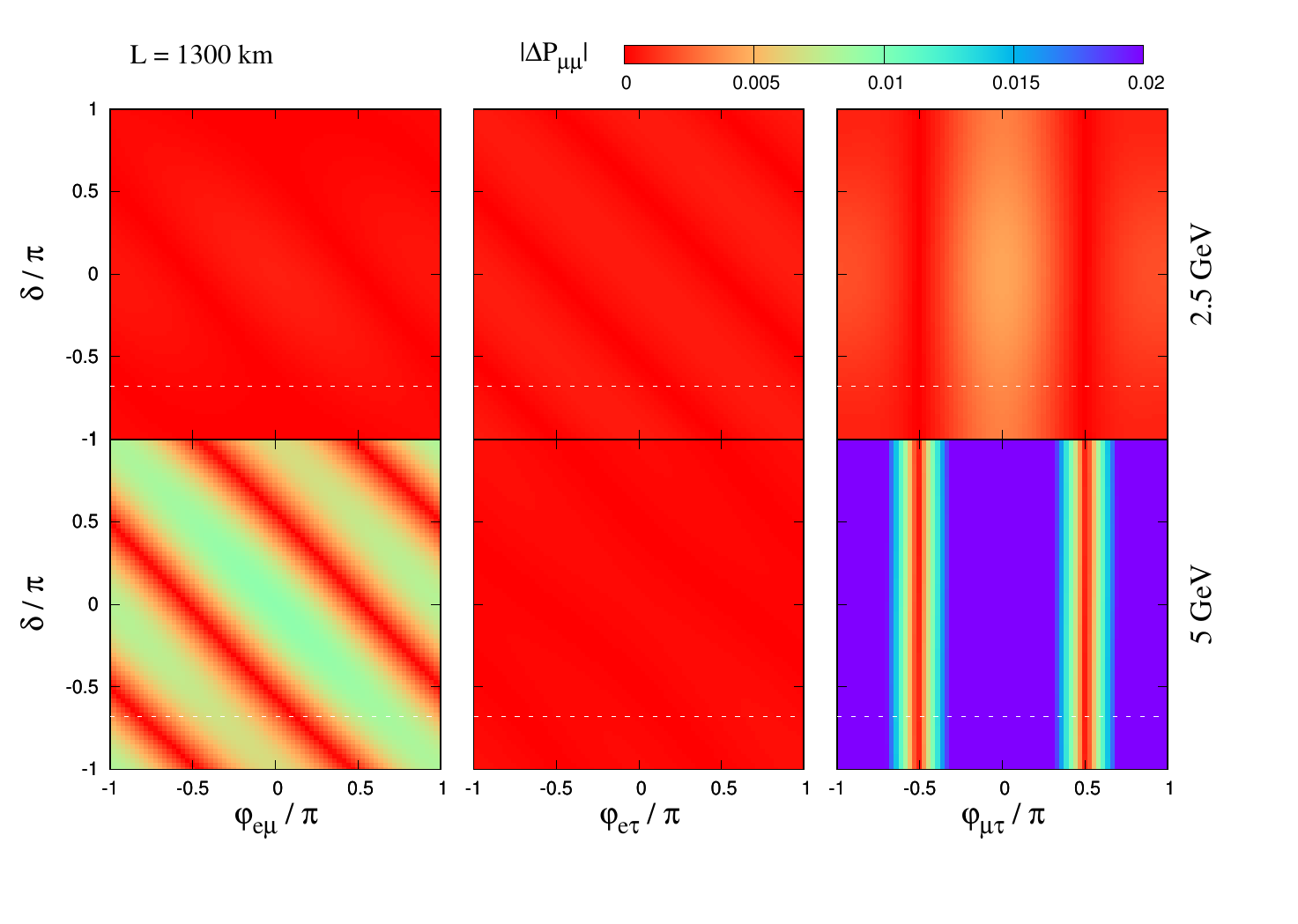}
\caption{\footnotesize{Similar to Fig.~\ref{fig:pme_phi_deg_del} but for the $\nu_{\mu} \rightarrow \nu_{\mu}$ channel.}}
\label{fig:pmm_phi_deg_del}
\end{figure}

\item In the case of $\nu_\mu \to \nu_e$ channel (Fig.~\ref{fig:pme_eps_deg_del}), the NSI terms $\varepsilon_{e\mu}$ and $\varepsilon_{e\tau}$ have relatively larger impact than the other NSI parameters.
For $\varepsilon_{e\mu}$ and $\varepsilon_{e\tau}$, the degenerate regions ($|\Delta \pme| \lesssim 0.05$) are narrowly concentrated around a pair of values of $\delta$ (see Table~\ref{tab:pme_em_et_location_num} below)
\begin{table}[ht]
\centering
\begin{tabular}{|c|c|c|}
\hline
E & $\Delta P_{\mu e}(\eema) \approx 0$ & $\Delta P_{\mu e}(\eeta) \approx 0$\\ \hline
2.5 GeV & $0.25\pi, -0.8\pi$ & $0.8\pi, -0.2\pi$\\ \hline
5 GeV & $0.4\pi, -0.6\pi$ & $0.95\pi, -0.15\pi$\\ \hline
\end{tabular}
\caption{\label{tab:pme_em_et_location_num}
The values of $\delta$ where $|\Delta \pme|$ almost vanishes in the presence of $\eem$ or $\eet$  ({red spikes}) in Fig.~\ref{fig:pme_eps_deg_del}.}
\end{table} 
These sharp SI-NSI degenerate regions exist even for 5 GeV but at somewhat different values of $\delta$. 
For $\eet$, the degenerate region seems to be larger at $5$ GeV in contrast to $2.5$ GeV. This is not seen in the case of $\eem$ (this observation is consistent with Fig.~\ref{fig:eps_deg_en}).
Note that the locations of SI-NSI degenerate regions is roughly independent of the size of $\eema$ and $\eeta$.
For  $|\varepsilon_{\mu\tau}| $, the degenerate region is broader and shows a soft feature of peaking at $\delta \approx \pm \pi$ for $2.5$ GeV. 
For  $\varepsilon_{ee} $ and $\varepsilon_{\tau\tau} $, the degenerate regions have similar structure showing no CP phase dependence.  

\item For the $\nu_{\mu} \to \nu_{\mu}$ channel (Fig.~\ref{fig:pmm_eps_deg_del}), as mentioned earlier,  it is more appropriate to look at $5$ GeV (the bottom row).
As expected, $\emt$ has the largest impact and its effect is independent of the CP phase, $\delta$ (see also Appendix~\ref{sec:pmm_analysis}). 
The impact of $\eem$ is also important with two sharp peaks occurring around $\delta \approx \pm \pi/2$. The other terms such as $\eet, \eee, \ett$ have almost no effect at $5$ GeV 
(here also the results are consistent with Fig.~\ref{fig:eps_deg_en}). 
\end{itemize}

To complete the discussion, we now discuss the effect of non-zero phases. 
We keep the moduli of the respective NSI terms fixed at $|\varepsilon_{\alpha\beta}| = 0.05$ and plot heatmaps corresponding to $|\Delta P_{\mu e}|$ and $|\Delta P_{\mu \mu}|$ in the $\varphi_{\alpha\beta} - \delta$ plane in Fig.~\ref{fig:pme_phi_deg_del} and Fig.~\ref{fig:pmm_phi_deg_del} respectively. 
As before, we show our results for two different values energy, $2.5$ GeV (top row) and $5$ GeV (bottom row).
 We make the following observations from these plots: 
\begin{itemize}
\item In the case of the $\nu_{\mu} \to \nu_{e}$ channel (Fig.~\ref{fig:pme_phi_deg_del}), 
we see degenerate regions  in the case of $\varphi_{e \mu}$ and $\varphi_{e \tau}$ (where $|\Delta \pme| \lesssim 0.005$) slanted at an angle of $135^{o}$.  
In the case of $\varphi_{\mu\tau}$,  $|\Delta \pme|$ remains close to zero and  stays within $\lsim 0.005$ in the entire $\varphi_{\mu\tau} - \delta$ space~\footnote{In general, $\epsilon_{\mu \tau}$ has milder impact on the $P_{\mu e}$. 
 The effect of the associated NSI phase $\emtp$ is, thus, small. 
 If we take somewhat larger value of $\emta$, $|\Delta \pme|$ would increase slightly but the qualitative feature of $|\Delta \pme (\emtp)|$ would still remain similar.}. 
For $5$ GeV, the pattern remains very similar for $\eemp$, but the extent of degeneracy increases for $\eetp$, as expected from our previous analyses. 

From the analytic expressions given in Appendix~\ref{sec:pme_analysis} 
(Eq.~\ref{eq:pme_eem} and \ref{eq:pme_eet}), we can note that 
the SI-NSI degeneracy in the presence of $\eemp$ or $\eetp$ for a fixed non-zero moduli of the corresponding NSI term, arises from the following:
\begin{align}
\label{eq:phi_deg_cond}
&\sin(\delta + \eemp - \gamma_{1}^{e\mu}) \approx 0 ~(\text{for} ~\eemp) \quad 
{\textrm{and}} \quad \sin(\delta + \eetp + \gamma_{1}^{e\tau}) \approx 0 ~(\text{for} ~\eetp) 
\nonumber \\
&\implies \delta + \eemp \approx n\pi + \gamma_{1}^{e\mu} \quad {\textrm{and}} \quad \delta + \eetp \approx n\pi - \gamma_{1}^{e\tau}~,  {\textrm{with~}} n = 0, \pm 1, \pm 2, \ldots 
\end{align}
Here $\gamma_{1}^{e\mu} = \tan^{-1}(\frac{\tan^{2}\theta_{23}}{\Delta} + \cot\Delta)$ and $\gamma_{1}^{e\tau} = \tan^{-1}(\frac{1}{\Delta} - \cot\Delta)$.
We note that Eqns.~\ref{eq:phi_deg_cond} show equations of straight lines with a slope of $135^{o}$ and equal intercepts on the $\delta$ and $\varphi_{\alpha\beta}$ axes~\footnote{$x/a + y/b = 1$ is a general equation of straight line with intercepts $a$ and $b$ on 
the $x$ and $y$ axes respectively.}.
Furthermore, the various intercepts (corresponding to different $n$) on the $\varphi_{\alpha\beta}$ or $\delta$ axes are separated by $\pi$ which is also seen in Fig.~\ref{fig:pme_phi_deg_del}.

\item In case of the $\nu_{\mu} \to \nu_{\mu}$ channel (Fig.~\ref{fig:pmm_phi_deg_del}), we focus on 
the bottom row. Here $\eemp$ shows the SI-NSI degenerate regions roughly mimicking straight lines at $135^{o}$ slope, whereas $\eetp$ shows no effect. 
$\emtp$ manifests itself by rendering $|\Delta \pmm|$ to a much larger value ($\gtrsim 0.02$) for most of the parameter space, but there exist two sharp degenerate regions occurring at $\emtp \approx \pm \pi/2$ with no $\delta$ dependence.

\end{itemize}

%----------------------------------------------------------

\section{Probing the NSI parameter space at the level of $\chi^2$}
\label{sec:chisq}

In the present section, we numerically explore the NSI parameter space at the level of $\chi^2$  using the standard LE as well as ME beam tunes. Our main results are summarized in Fig.~\ref{fig:contour} where we depict contours at a confidence level (c.f.) of $99\%$. 
The solid cyan (black hatched) contours correspond to LE (LE + ME) beams. 
More specifically, the regions enclosed by these contours depict the regions where there is SI-NSI degeneracy for those pair of parameters. Below, we discuss some noteworthy features as can be observed from
 Fig.~\ref{fig:contour}:

%----------------------------------------------------------

\begin{figure*}[h]
\centering
\includegraphics[width=1\textwidth]
{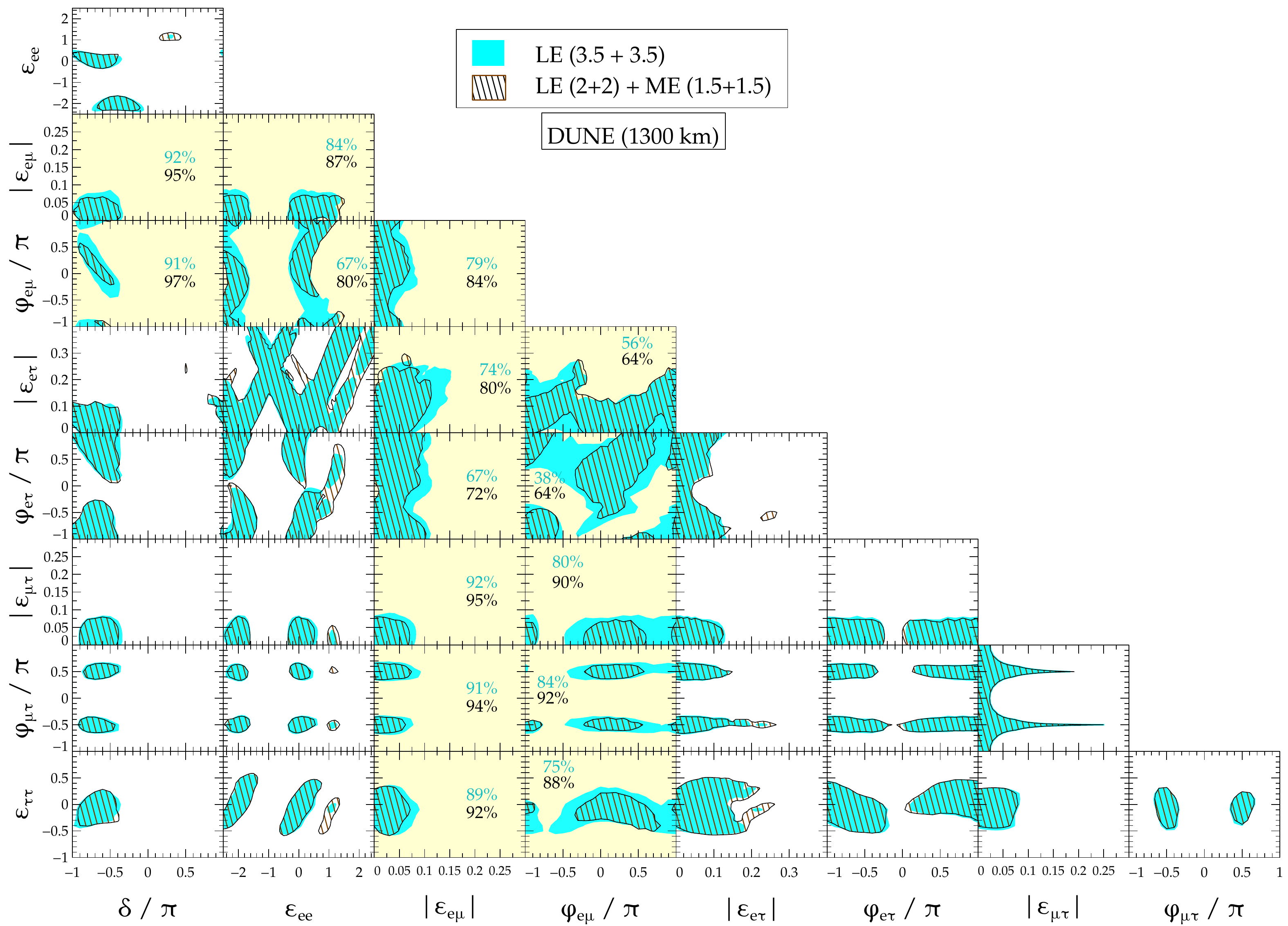}
\caption{\footnotesize{A comparison of the sensitivity of DUNE to probe the NSI parameters at $99\%$ confidence level 
when a standard low energy (LE) beam tune is used (cyan region) and when a  combination of low and medium energy (LE + ME) beam tune is used (black hatched region), keeping the total runtime same (3.5 years of $\nu$ + 3.5 years of $\bar{\nu}$ run) for both scenarios. In the latter case, the total runtime is distributed between the LE beam (2 years of $\nu$ + 2 years of $\bar{\nu}$) and the medium energy beam (1.5 years of $\nu$ + 1.5 years of $\bar{\nu}$).
The panels with a light yellow (white) background indicate significant improvement (no improvement) by using LE + ME beam over using LE only.  The numbers in the light yellow shaded panels correspond to the area lying outside the contour for the two cases (cyan for LE and black for LE+ME) expressed as a percentage of the total parameter space plotted. These numbers quantify the improvement over the LE only option when the ME beam tune is used in conjunction with the LE beam tune in these panels.
 }}
\label{fig:contour}
\end{figure*}

%----------------------------------------------------------

\begin{enumerate}
\item Let us first consider the panels with $\eem$ (either $\eema$ 
or $\eemp$ or both) which are shown in light yellow colour. 
We note that use of different beam tunes (ME in conjunction with the LE beam) offers visible improvement of results (shrinking of contours) in these pairs of parameters. This is one of the key results of the present article.  
In order to explain the observed pattern, let us recollect from 
 Figs.~\ref{fig:eps_deg_en} and \ref{fig:phi_deg_en} that the presence of $|\eem|$ or $\eemp$ leads to large difference between SI and NSI scenarios even at larger values of energies \ie, $E \gtrsim 4$ GeV. 
Thus, with the LE+ME option we are able to place tighter constraints on the parameter space corresponding to parameters $|\eem|$ and $\eemp$. 

From Eq.~\ref{eq:chisq_si_nsi}, the $\Delta \chi^{2}$ in the presence of two NSI parameters, say, $a$ and $b$, can  be written as~\footnote{For the ease of understanding, we write neutrino contribution only. The dependence on flux and cross-section has been omitted for clarity in understanding the dependence on probabilities.}:
\begin{align}\label{eq:chisq_mech}
&\Delta \chi^{2} (a,b) \sim \Delta \chi^{2}_{\mu e} (a,b) + \Delta \chi^{2}_{\mu\mu} (a,b) \nonumber \\
& \sim \textrm{Min}\sum_{\textrm{energy}} \bigg[ |\Delta \pme (a)| + |\Delta \pme (b)| + |\Delta \pmm (a)| + |\Delta \pmm (b)| \bigg].
\end{align}
For \eg, if we focus on the $\eema$-$\eeta$ plane, we have%
\begin{align}\label{eq:chisq_mech_eem_eet}
\Delta \chi^{2} (\eema, \eeta) \sim \text{Min}\sum_{\text{energy}} \bigg[ 
|\Delta \pme (\eema)| + |\Delta \pme (\eeta)| 
+ |\Delta \pmm (\eema)| + |\Delta \pmm (\eeta)| 
\bigg],
\end{align}
where the sum is over all the energy bins ($0-20$ GeV) and the minimization is performed over 
$\delta, \theta_{23}, \Delta m^{2}_{31}, \eemp, \eetp$. 
From the probability level discussion (Fig.~\ref{fig:eps_deg_en}), we can assess the impact of the NSI terms $\eema$ and $\eeta$ on $\pme$ and $\pmm$. In the case of $|\Delta \pme|$, at low values of energy, the impact of the two NSI parameters is quite similar. But, at higher energies, the effects due to $\eema$  tend to be larger than effects  due to $\eeta$. This means that ME beam is expected to alter the degenerate region more in the case of $\eema$ and less in the case of $\eeta$. That the smaller contribution from the disappearance channel is in the same direction as the larger contribution from the appearance channel (with $\eema$ and $\eeta$ acting in opposite directions) can also be seen from the plot. 

\item 
We next consider the remaining panels in which we see that there is 
very little or no improvement of results after using the ME beam along with the LE beam. 
If we look at the pair of parameters, $\eeta-\eee, \eetp-\eee, \ett-\eee, \eetp-\eeta$ and $\ett-\eeta$ in particular, we note that the degenerate regions get enlarged slightly. 
This is because of the fact that the the presence of $\eet$, unlike $\eem$, actually adds to the SI-NSI degeneracy at higher energies (see Figs.~\ref{fig:eps_deg_en} and \ref{fig:phi_deg_en} and the discussions in Sec.~\ref{sec:prob_heatmap_energy}). 

\item For the panels with $\emta$ and $\emtp$ as one of the parameters, there is very  marginal improvement (except when $\eema$ or $\eemp$ is present) in the degenerate contours using the LE+ME beam. 
To see how the $\Delta \chi^{2}$ arises in panels showing the parameter space associated with $\emta$, let us take for example, the pair of parameters, $\emta$ and $\eeta$ and express the $\Delta \chi^2$ (Eq.~\ref{eq:chisq_mech}) as  
\begin{align}
\label{eq:chisq_mech_emt_eet}
\Delta \chi^{2} (\emta, \eeta) \sim \text{Min}\sum_{\text{energy}} \bigg[ |\Delta \pme (\emta)| + |\Delta \pme (\eeta)| 
+ |\Delta \pmm (\emta)| + |\Delta \pmm (\eeta)| 
\bigg],
\end{align}
where the sum is over all the energy bins ($0-20$ GeV) and the minimization is carried over 
$\delta, \theta_{23}, \Delta m^{2}_{31}, \emtp, \eetp$. 
Now, from Eq.~\ref{eq:pmm_emt}, we know that in leading order, $|\Delta \pmm(\emt)|$ is independent of $\delta$ and is directly proportional to $\cos\emtp$. 
Minimization over $\emtp \in [-\pi, \pi]$ will always then find the constant, energy-independent value of $\emtp \approx \pm \pi/2$ which makes the $\Delta \chi^{2}$ contribution due to $\pmm$ 
 vanishingly small%
\footnote{On the other hand this does not happen for $\eem$ and $\eet$ for the following reason. Eqns.~\ref{eq:pme_eem} and \ref{eq:pme_eet} tell us that in leading order, $|\Delta \pme(\eem)| \propto \sin(\delta + \eemp - \gamma_{1}^{e\mu})$ and $|\Delta \pme(\eet)| \propto \sin(\delta + \eetp + \gamma_{1}^{e\tau})$ where $\gamma_{1}^{e\mu}$ and $\gamma_{1}^{e\tau}$ are energy-dependent quantities. 
Thus, unlike in the case of $|\Delta \pmm (\emt)|$, there does not exist a unique energy-independent phase value which would make its contribution to $\Delta \chi^{2}$ to $\sim 0$.}. 
Thus, even when $\emta$ is present, the $\Delta \chi^{2}$ receives a dominant contribution from the $\nu_{\mu} \to \nu_{e}$ channel. 
This is more clear from the panels showing the parameter space associated to $\emtp$ (\ie, where $\emtp$ is not marginalised). 
The magnitude of $\Delta \chi^{2}$ in such panels is dominantly contributed by the $\nu_{\mu} \to \nu_{\mu}$ channel for all values of $\emtp \not\approx \pm \pi/2$. 
But around $\emtp \approx \pm \pi/2$, the contribution from the $\nu_{\mu} \to \nu_{\mu}$ becomes very small and the $\nu_{\mu} \to \nu_{e}$ channel dominates, as we have also verified numerically. 
This explains the appearance of degenerate contours at $\emtp \approx \pm \pi/2$ as well.

\item
All the parameter spaces showing $\eee$ (entire 2nd column and the top panel of the 1st column) have an additional degeneracy around $\eee \approx -2$, in addition to the true 
solution at $\eee \approx 0$. This extra solution comes due to the marginalisation over the 
opposite {{mass hierarchy}}. Similar degeneracy has also been observed in previous studies: in
\cite{Miranda:2004nb, Coloma:2016gei, Deepthi:2016erc} (in the context of NSI) and also in \cite{Barenboim:2018ctx} in the context of Lorentz violating parameters.

\end{enumerate}

\section{Concluding remarks}\label{sec:conclude}

 In ~\cite{Masud:2017bcf}, the authors combined the experimentally feasible option  of using a  combination of beam tunes and demonstrated that it was possible to extricate any two physics scenarios more efficiently using experimental handles. In the present article, we address the question of constraining the parameter space of NSI parameters at DUNE by exploiting wide band nature of the beam. We systematically study correlations among various parameters  using two beam tunes (LE and ME) and illustrated that to probe a subset of NSI parameter space more effectively, it is advantageous to use a combination of LE and ME tuned beams as opposed to using only the standard LE beam tune. 

We provide a systematic and  comprehensive description of the overall impact of the NSI parameters on the relevant probabilities (for $\nu_\mu \to \nu_e$ and for $\nu_\mu\to \nu_\mu$) as a function of energy as well as the CP phase. In the Appendices, we provide analytic expressions of all the relevant expressions for the  SI-NSI probability differences  in the presence of individual NSI parameters (taken one at a time). These aid in our understanding of the dependencies of oscillation probabilities.  We then quantify the differences in terms of a $\Delta \chi^2$ quantity and connected the features obtained to the probability level description. In Fig.~\ref{fig:contour}, we have illustrated the $\Delta \chi^{2}$ correlations among the various parameters in the new parameter space appearing in the presence of NSI at a confidence level of $99\%$. Our key findings can be summarized as follows. The degenerate contours in the space associated with parameters, $\eema$ and $\eemp$ (shown as panels shaded in light yellow colour in Fig.~\ref{fig:contour}) shrink significantly when we use the LE+ME beam as opposed to LE beam alone. For a quantitative estimate of the improvement,  one can compute the  area of the parameter space outside each contour (i.e., above the confidence level of $99\%$) and express the area as the percentage of the total parameter space plotted. 
  It is evident from the pair of numbers (cyan for LE and black for LE+ME) indicated in the light yellow panels that the LE+ME beam leads to improvement over the LE beam.
 For the remaining NSI parameters, we see marginal or no improvement in terms of constraining the parameters using LE+ME beam in comparison with LE beam. Our detailed analysis also provides explanation for 
distinguishing features of the $\Delta \chi^{2}$ contours for different parameters.

A few remarks on connection with the existing work that deal with constraining NSI parameters at DUNE~\cite{deGouvea:2015ndi, Coloma:2015kiu} are in order. It should be noted that  the standard beam (available in 2015) was used for these analyses.  
It can be observed that the contours in our analyses indicate better resolution capability of DUNE and they roughly resemble the contours of 
~\cite{deGouvea:2015ndi, Coloma:2015kiu} in shape. The slight differences in the contours may arise from the difference in the experimental inputs such as beam configuration, detector details, exposure and the best-fit values used.

Although the entire analysis has been carried out in the context of NSI and DUNE, we would like to mention that the approach is fairly general and can be easily translated to other new physics contexts such as non-unitarity, CPT violation,  Lorentz violation etc. We point out that if we utilize the full 
 potential of a given long baseline experiment such as DUNE which has wide band beam and allows for 
 tunability of beam, we can reduce the degeneracy (using the same experiment with multiple beam tune options) among some relevant choice of
  parameters in the parameter space as is suggested by the probability level discussion 
  in Sec.~\ref{sec:prob}.

%~~~~~~~~~~~~~~~~~~~~~~~~~~~~~~~~~~~~~~~~~~~~~

%\section{Appendices}
\appendix
\setcounter{equation}{0}
\setcounter{section}{0}
\renewcommand{\thesection}{\Alph{section}}
\renewcommand{\theequation}{\thesection\arabic{equation}}
\section{Analytic understanding of the behaviour of $\Delta \pme$} 
%-----------------------------
\label{sec:pme_analysis}
%-----------------------------
Here we look at the expressions for probability difference between SI and NSI  and make an attempt in understanding how the individual NSI parameters affect the SI-NSI degeneracy. We calculate these expressions by making use of the probability expressions from \cite{Kikuchi:2008vq} upto first order in $\varepsilon_{\alpha\beta}$'s. Using the expressions for $\pme$ in presence 
of a single NSI parameter ($\eema$, $\eeta$ or $\eee$) we arrive at the following three equations:

\begin{align}\label{eq:pme_eem}
&\Delta \pme(\eem) = P_{\mu e}^{NSI}(\eem) - P_{\mu e}^{SI} \nonumber \\
%%%%%%%%%%%%%%
 &\underbrace{\approx -4A\Delta \sin \Delta \eema s_{13} s_{2(23)} c_{23} D_{1}^{e\mu} \sin(\delta + \eemp - \gamma_{1}^{e\mu})}_{\textrm{\footnotesize{\textcolor{green}{I}}}} \quad
 + \quad \underbrace{2A\Delta \sin \Delta \eema \alpha s_{2(12)} s_{2(23)} s_{23} D_{2}^{e\mu} \sin(\eemp + \gamma_{2}^{e\mu})}_{\textrm{\footnotesize{\textcolor{blue}{II}}}} \quad + \quad O(\eem^{2})
 \nonumber \\
 &\approx 2A\Delta \sin \Delta \eema s_{2(23)}\bigg[
 -2s_{13} c_{23} D_{1}^{e\mu} \sin(\delta + \eemp - \gamma_{1}^{e\mu}) 
 + \alpha s_{2(12)} s_{23} D_{2}^{e\mu} \sin(\eemp + \gamma_{2}^{e\mu})
 \bigg],
\end{align}
\&
\begin{align}\label{eq:pme_eet}
&\Delta \pme(\eet) = P_{\mu e}^{NSI}(\eet) - P_{\mu e}^{SI} \nonumber \\
%%%%%%%%%%%%%%
 &\underbrace{\approx 4A\Delta \sin \Delta \eeta s_{13} s_{2(23)} s_{23} D_{1}^{e\tau} \sin(\delta + \eetp + \gamma_{1}^{e\tau})}_{\textrm{\footnotesize{\textcolor{green}{I}}}} \quad
 + \quad \underbrace{\big(-2A\Delta \sin \Delta \eeta \alpha s_{2(12)} s_{2(23)} c_{23} D_{2}^{e\tau} \sin(\gamma_{2}^{e\tau} - \eetp)\big)}_{\textrm{\footnotesize{\textcolor{blue}{II}}}} \quad + \quad O(\eet^{2})
 \nonumber \\
 &\approx 2A\Delta \sin \Delta \eeta s_{2(23)}\bigg[
 2s_{13} s_{23} D_{1}^{e\tau} \sin(\delta + \eetp + \gamma_{1}^{e\tau}) 
 - \alpha s_{2(12)} c_{23} D_{2}^{e\tau} \sin(\gamma_{2}^{e\tau} - \eetp)
 \bigg],
\end{align}
where,
\begin{align} 
&D_{1}^{e\mu} = [\sin^{2}\Delta + (\tan^{2}\theta_{23}\frac{\sin \Delta}{\Delta} + \cos \Delta)^{2}]^{1/2}  
&\gamma_{1}^{e\mu} = \tan^{-1}(\frac{\tan^{2}\theta_{23}}{\Delta} + \cot\Delta)
\nonumber \\
&D_{2}^{e\mu} = [\sin^{2}\Delta + (\cot^{2}\theta_{23}\frac{\Delta}{\sin \Delta} + \cos \Delta)^{2}]^{1/2} 
&\gamma_{2}^{e\mu} = \tan^{-1}(\frac{\cot^{2}\theta_{23} \Delta}{\sin^{2}\Delta} +\cot\Delta) \nonumber \\
&D_{1}^{e\tau} = [\sin^{2}\Delta + (\frac{\sin \Delta}{\Delta} - \cos \Delta)^{2}]^{1/2} ;
&\gamma_{1}^{e\tau} = \tan^{-1}(\frac{1}{\Delta} - \cot\Delta) \nonumber \\
&D_{2}^{e\tau} = [\sin^{2}\Delta + (\frac{\Delta}{\sin \Delta} - \cos \Delta)^{2}]^{1/2} 
&\gamma_{2}^{e\tau} = \tan^{-1}(\frac{\Delta}{\sin^{2}\Delta} - \cot\Delta). \nonumber
\end{align}
Here $A = a/\Delta m^2_{31} = 2 \sqrt{2}  E G_F n_e / \Delta m^2_{31}$. 
By making the substitution $A \to A(1+\eee)$~\cite{Liao:2016hsa} we also have,
\begin{align}\label{eq:pme_eee}
& \Delta P_{\mu e} (\eee) = P_{\mu e}^{NSI}(\eee) - P_{\mu e}^{SI} \nonumber \\
&\approx {\underbrace{
s_{2(13)}^{2}s^{2}_{23}\bigg[
\frac{\sin^{2}\big[\big\{1-A(1+\eee) \big\}\Delta \big]}{\big\{1-A(1+\eee) \big\}^{2}}
- \frac{\sin^{2}\big\{\big(1-A \big)\Delta \big\}}{\big(1-A \big)^{2}}
\bigg] 
%\text{\footnotesize{\textcolor{green}{$\leftarrow$ Term I }}} 
}_{\textcolor{green}
 {\textrm{I}}}}
\nonumber \\
%%%%%%%%%%%%%%%%%
&  % {\underbrace{ 
+ \quad \Bigg\{
\alpha^{2}s^{2}_{2(12)}c^{2}_{23}
\bigg[
\frac{\sin^{2}\big\{A(1+\eee) \Delta \big\}}{\big\{A(1+\eee) \big\}^{2}}
- \frac{\sin^{2}\big(A \Delta \big)}{A^{2}}
\bigg] + 
%}}
%\text{\footnotesize{\textcolor{blue}{$\leftarrow$ Term II }}} 
\nonumber \\
%%%%%%%%%%%%%%%%%%%%
&  \quad \quad {\underbrace{  \quad
\alpha s_{2(13)}s_{2(12)}s_{2(23)} \bigg[
\frac{\sin\big[\big\{1-A(1+\eee) \big\}\Delta \big]}{1-A(1+\eee)}
.\frac{\sin\big\{A(1+\eee) \Delta \big\}}{A(1+\eee) }
- \frac{\sin\big\{\big(1-A \big)\Delta \big\}}{1-A }
.\frac{\sin\big(A \Delta \big)}{A }
\bigg]\cos(\delta + \Delta) \Bigg\}
%\text{\footnotesize{\textcolor{cyan}{$\leftarrow$ Term III }}}
}_{\textcolor{blue}
 {\textrm{II}}}}~,
\end{align}
where $\Delta = \frac{\ldm L}{4E}$.

\begin{figure*}[tb]
\centering
\includegraphics[scale=0.6]{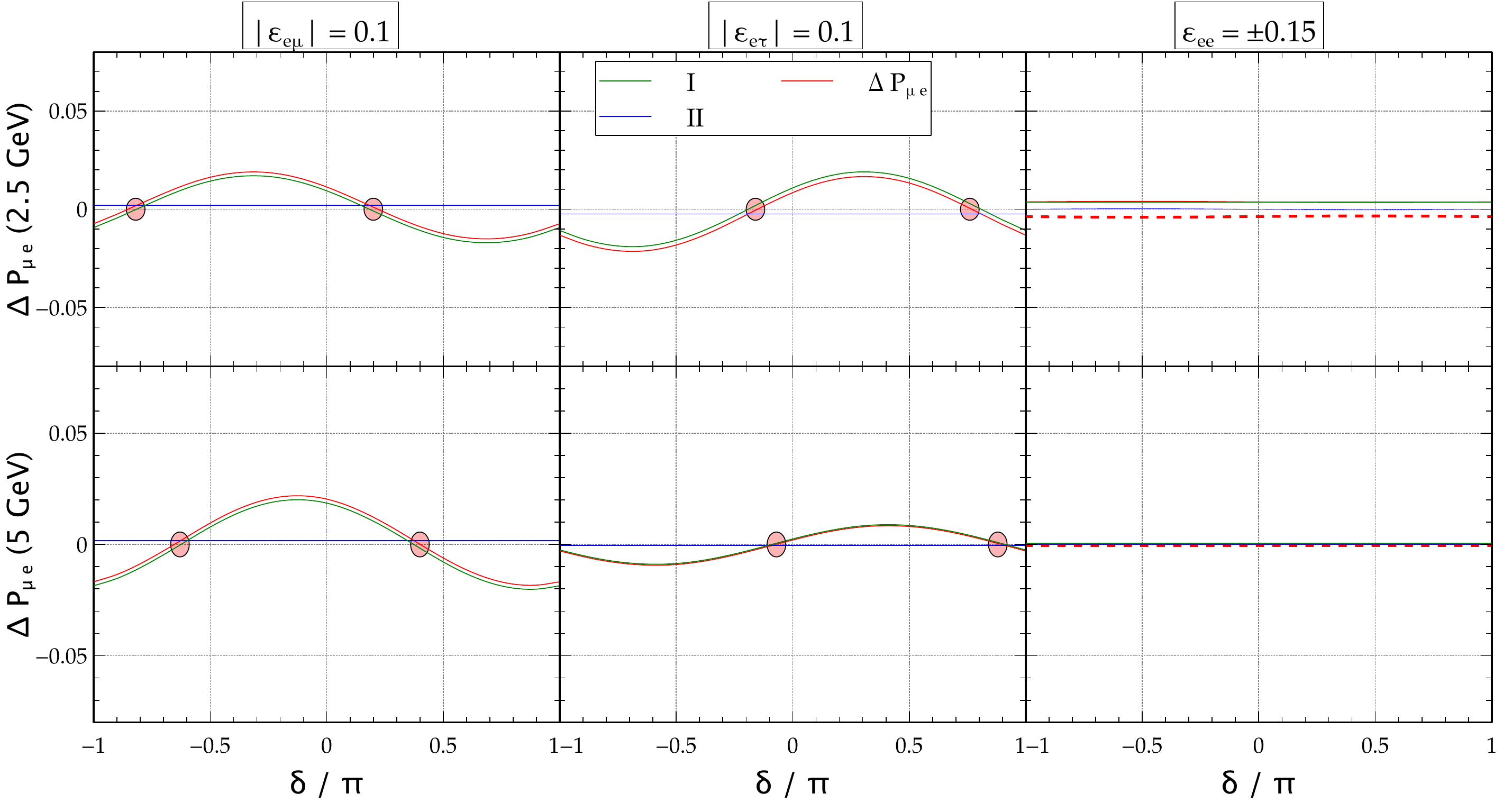}
\caption{\footnotesize{The  terms (denoted by green, blue and cyan curves) in the RHS of Eq.~\ref{eq:pme_eem} (first column), \ref{eq:pme_eet} (second column) and \ref{eq:pme_eee} (third column) are plotted as functions of $\delta$ for two fixed energies $2.5$ GeV (top row) and $5$ GeV (bottom row). The overall $\Delta \pme$ is represented by the red curve and the small red circles denote where it becomes zero. 
}}
\label{fig:deg_pme}
\end{figure*}

When $\Delta \pme$ becomes close to zero, it becomes difficult to separate NSI from SI and we have a SI-NSI degeneracy. 
We plot the terms in Eq.~\ref{eq:pme_eem}, Eq.~\ref{eq:pme_eet} and Eq.~\ref{eq:pme_eee} as functions of $\delta$ for an energy of 2.5 GeV and also at a higher energy of 5 GeV in  Fig.~\ref{fig:deg_pme} with fixed values of the NSI amplitude and zero NSI phase as indicated in the figure.
For $\eem$ or $\eet$, the second term (blue) is very small (scaled down by the additional factor $\alpha \sim 10^{-2}$ compared to the first term) and also independent of the CP phase $\delta$. 
It is the first term (green) which mainly dictates the behaviour of $\Delta \pme$ in presence of $\eem$ or $\eet$. 
We note the locations (see Table \ref{tab:pme_em_et_location}) where the overall value of $\Delta \pme$ (red) 
becomes zero in Fig.~\ref{fig:deg_pme}. 
These locations are indeed consistent with the locations of the \textit{red spikes} in Fig.~\ref{fig:pme_eps_deg_del} 
 as listed in Table \ref{tab:pme_em_et_location_num}.
 The origin of these special values of $\delta$ can easily be understood as follows.

\begin{figure*}[htb]
\centering
\includegraphics[scale=0.6]{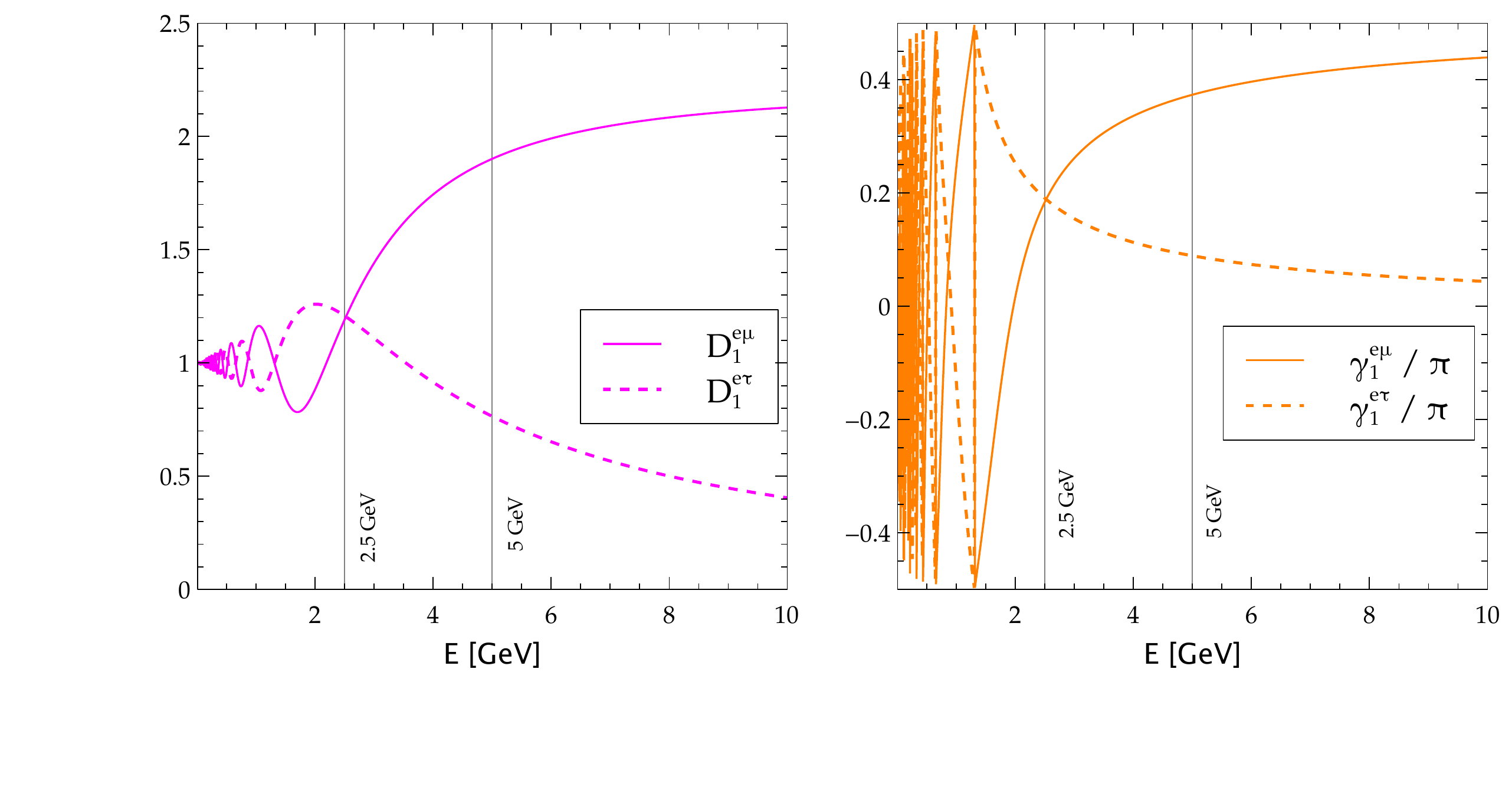}
\caption{\footnotesize{$D_{1}^{e\mu}$ and $D_{1}^{e\tau}$ are plotted as functions of energy (left panel).
 $\gamma_{1}^{e\mu}$ and $\gamma_{1}^{e\tau}$ are plotted as functions of energy in the right panel.
 The standard oscillation parameters are at their b.f value (Table \ref{tab:parameters}).
}}
\label{fig:em_et_en_dep}
\end{figure*}

On a closer inspection of the first term in Eq.~\ref{eq:pme_eem} and Eq.~\ref{eq:pme_eet}, we observe that 
it is proportional to $D_{1}^{e\mu}$ for $\eem$ and to $D_{1}^{e\tau}$ for $\eet$. 
From Fig.~\ref{fig:em_et_en_dep} (left panel), we observe that around 2.5 GeV 
both $D_{1}^{e\mu}$ and $D_{1}^{e\tau}$ have similar magnitude~\footnote{Recall that the b.f. value of $\theta_{23}$ in our analysis is not maximal, rather $47.7^{o}$. Even then the octant does not appear to play a significant role despite the presence of the extra $\tan^{2}\theta_{23}$ factor in the definition of $D_{1}^{e\mu}$.}. 
But as the energy increases further  $D_{1}^{e\mu}$ keeps on increasing while 
$D_{1}^{e\tau}$ decreases. 
This indicates that at higher energies, $|\Delta \pme (|\eem|)|$ increases while $|\Delta \pme (|\eet|)|$ becomes smaller. 
This explains why the degeneracy increases for higher energy in presence of $\eet$ compared to $\eem$ in the 
$\nu_{e}$ appearance channel, as observed from our simulation  earlier(see Fig.~\ref{fig:eps_deg_en}).

On a related note, let us also try to understand the role of $\varphi_{\alpha\beta}$ in Fig.~\ref{fig:phi_deg_en} 
with variation in energy. 
In Fig.~\ref{fig:em_et_en_dep} (right panel) we show the variation of the phase arguments $\gamma_{1}^{e\mu}$ and $\gamma_{1}^{e\tau}$ (appearing in the first terms of Eq.~\ref{eq:pme_eem} and Eq.~\ref{eq:pme_eet}) with energy. 
Around 2.5 GeV, $\gamma_{1}^{e\mu} \approx \gamma_{1}^{e\tau}$. At higher energies,  $\gamma_{1}^{e\mu} > \gamma_{1}^{e\tau}$, but both remains positive. 
Both of them tends to get plateaued at $E \gtrsim 4$ GeV or so, with $ \gamma_{1}^{e\mu} / \pi \sim 0.3$ 
and $\gamma_{1}^{e\tau} /\pi \sim 0.1$ on an average to a crude approximation.
Since $|\Delta \pme (\eem)| \propto \sin(\delta + \eemp - \gamma_{1}^{e\mu})$ approximately, 
we can guess that with a given b.f. value of $\delta \approx -0.7\pi$ we will have a degeneracy around 
$\eemp /\pi \approx 0, \pm \pi$  at energies $\gtrsim 4$ GeV. 
Similarly, since $|\Delta \pme (\eet)| \propto \sin(\delta + \eetp + \gamma_{1}^{e\tau})$ approximately, 
 we will have a degeneracy around 
$\eemp /\pi \approx 0.6, -0.4$ at energies $\gtrsim 4$ GeV.  
A look at Fig.~\ref{fig:phi_deg_en} (top row: first and second columns) indeed shows that the 
heatmaps for $|\Delta \pme|$ looks similar around 2.5 GeV and at energies $\gtrsim 4$ GeV, the degenerate regions 
(red bands) become independent of energy and are located at the $\eemp$ (or $\eetp$) values just predicted above.

\begin{table}[ht]
\centering
\begin{tabular}{|c|c|c|}
\hline
E & $\Delta P_{\mu e}(\eema) \approx 0$ & $\Delta P_{\mu e}(\eeta) \approx 0$\\ \hline
2.5 GeV & $0.22\pi, -0.82\pi$ & $0.76\pi, -0.16\pi$\\ \hline
5 GeV & $0.4\pi, -0.63\pi$ & $0.92\pi, -0.1\pi$\\ \hline
\end{tabular}
\caption{\label{tab:pme_em_et_location}
The values of $\delta$ (obtained from Fig.\ \ref{fig:deg_pme}) where $\Delta \pme(|\eem|)$ and $\Delta \pme(|\eet|)$ (the red curves 
in Fig.\ \ref{fig:deg_pme}) becomes zero, giving rise to SI-NSI degeneracy.}
\end{table}
%---------------------------------------

%%%%%%%%%%%%%%%%%%%%%%%%%%%%%%

%We further discuss what happens if the 
%factor $\frac{\sin(1-A)\Delta}{(1-A)}$ of eq.\ \ref{eq:deg_pme} is $0$. In that case the NSI-SI degeneracy in fig.\ \ref{fig:pme_eps} is independent of $\delta$ (or in other words, the $1\%, 5\%$ contours in fig.\ \ref{fig:pme_eps} would become flat as a function of $\delta$). 

%%%%%%%%%%%%%%%%%%%%%%%%%%%%%%%%%%%%%%%%

\section{Probability analysis for $P_{\mu \mu}$}
\label{sec:pmm_analysis}

Proceeding along similar lines as Appendix~\ref{sec:pme_analysis}, we 
derive the expressions for $P_{\mu \mu}^{{NSI}} - P_{\mu \mu}^{{SI}}$. 
\begin{align}\label{eq:pmm_eem}
& \Delta P_{\mu\mu}(\eema) 
= P_{\mu\mu}^{NSI}(\eema) - P_{\mu\mu}^{SI}  \nonumber \\
& \approx  
  {\underbrace{-4s_{23}^{3}\frac{A}{1-A} \eema \bigg[A s_{23} \eema + 2s_{13}\cos\delta \bigg] X}_{\textcolor{green}
 {\textrm{I}}}}
 \nonumber \\
& {\underbrace{+ \quad 4s_{2(23)} \frac{Y}{A(1-A)}   
\bigg[ \alpha s_{13}s_{2(12)}\cos \delta  
  - (\alpha s_{2(12)} + Ac_{23}\eema  ) D(\eema) \cos(\delta - \theta (\eema))
\bigg]}_{\textcolor{blue}
 {\textrm{II}}}}
\end{align}

\begin{align}\label{eq:pmm_eet}
& \Delta P_{\mu\mu}(\eeta) 
= P_{\mu\mu}^{NSI}(\eeta) - P_{\mu\mu}^{SI}  \nonumber \\
& \approx  
  {\underbrace{-4s_{23}^{2} c_{23} \frac{A}{1-A} \eeta \bigg[A c_{23} \eeta + 2s_{13}\cos\delta \bigg] X}_{\textcolor{green}
 {\textrm{I}}}}
 \nonumber \\
& {\underbrace{+ \quad 4s_{2(23)} \frac{Y}{A(1-A)}    
\bigg[ \alpha s_{13}s_{2(12)}\cos \delta  
  - (\alpha s_{2(12)} - As_{23}\eeta ) D(\eeta) \cos(\delta - \theta (\eeta))
\bigg]}_{\textcolor{blue}
 {\textrm{II}}}}
\end{align}
where,
\begin{align} 
&X = c_{23}^{2}\Delta \sin 2\Delta
 + \frac{\sin^{2}(1-A)\Delta}{(1-A)}
- 2c_{23}^{2} \cos A\Delta \sin \Delta \frac{\sin(1-A)\Delta}{(1-A)} 
 \nonumber \\
&Y = c_{23}^{2}\sin^{2}A\Delta + s_{23}^{2}\sin^{2}(1-A)\Delta - s_{23}^{2}\sin^{2}\Delta 
- c_{2(23)} A \sin^{2}\Delta \nonumber \\
&D (|\eem|) = \bigg\{
 s_{13}^{2} + A^{2}s_{23}^{2}|\eem|^{2} + 2As_{13}s_{23}|\eem| \cos \delta
\bigg\}^{1/2} \quad ; \quad
\theta (\eema) = \text{arctan} \frac{As_{23}\eema \sin \delta}{s_{13} + As_{23}\eema \cos \delta}
 \nonumber \\
&D (|\eet|) = \bigg\{
 s_{13}^{2} + A^{2}c_{23}^{2}|\eet|^{2} + 2As_{13}c_{23}|\eet| \cos \delta
\bigg\}^{1/2} \quad ; \quad 
\theta (\eeta) = \text{arctan} \frac{Ac_{23}\eeta \sin \delta}{s_{13} + Ac_{23}\eeta \cos \delta}. \nonumber
\end{align}

\begin{figure}[ht]
\centering
\includegraphics[scale=0.5]{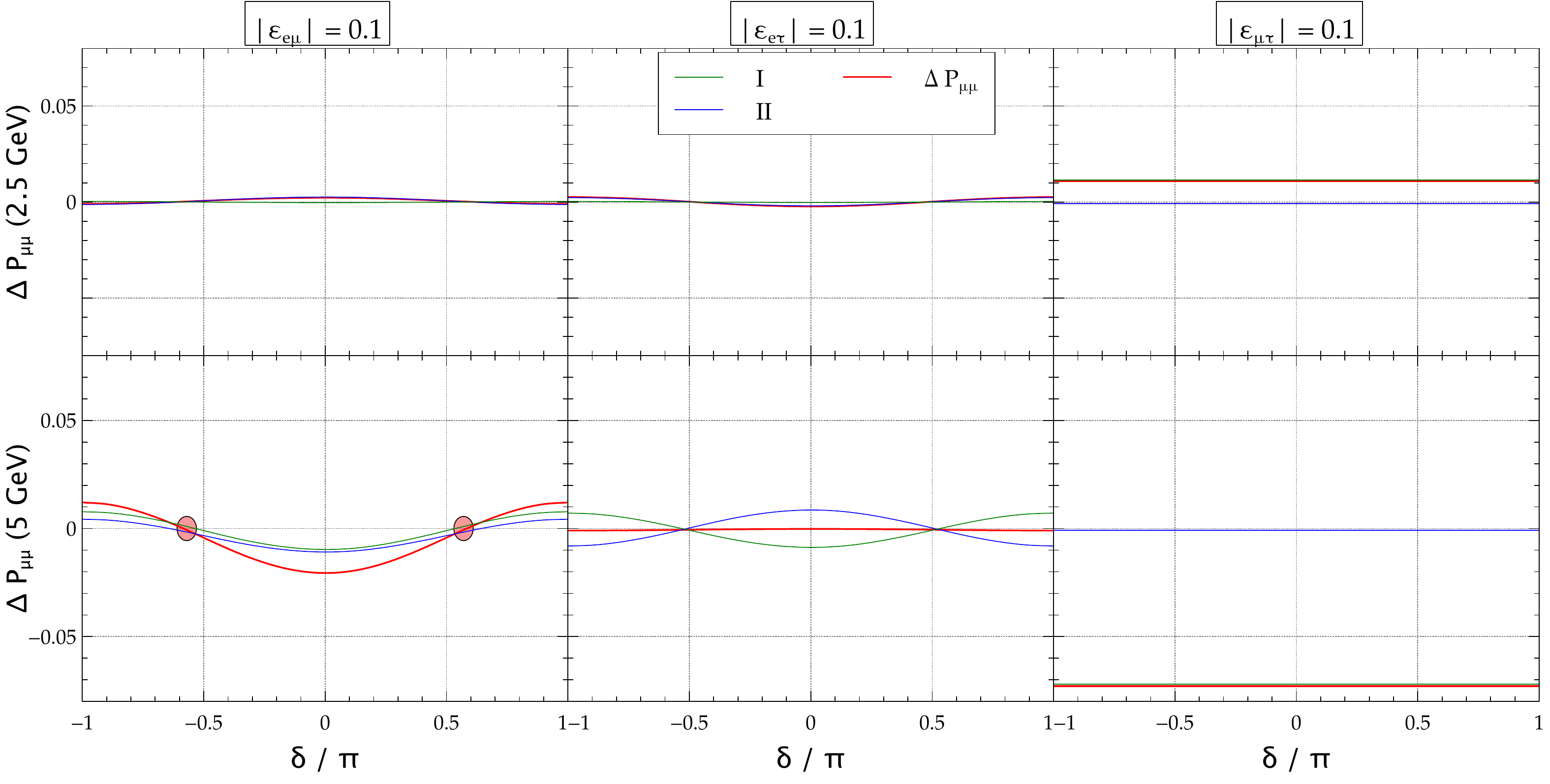}
\caption{\footnotesize{The terms (denoted by green, blue and cyan curves) in the RHS of Eq.~\ref{eq:pmm_eem} (first column), \ref{eq:pmm_eet} (second column) and \ref{eq:pmm_emt} (third column) are plotted as functions of $\delta$ for two fixed energies $2.5$ GeV (top row) and $5$ GeV (bottom row).  
The overall $\Delta \pmm$ (sum of the three terms) is represented by the red curve and the small red circles denote where it becomes zero.}}%
\label{fig:deg_pmm}
\end{figure}%

\begin{align}
& \Delta P_{\mu\mu}(\emt)  
 \approx 
{\underbrace{
(-2 \emta  A \Delta s_{2(23)}^{3} \sin 2\Delta \cos \emtp)}
_{\textcolor{green}
 {\textrm{I}}}} \quad 
+ \quad  {\underbrace{(
-4A \emta c^{2}_{2(23)} s_{2(23)} \sin^{2}\Delta \cos\emtp)}
_{\textcolor{blue}
 {\textrm{II}}}}
 \nonumber \\
 & 
\approx -4 \emta A s_{2(23)} \sin\Delta \cos \emtp 
[ \Delta s^{2}_{2(23)} \cos\Delta + c^{2}_{2(23)} \sin\Delta ].
\label{eq:pmm_emt}
\end{align}%

In Fig.~\ref{fig:deg_pmm}, we plot the terms of Eq.~\ref{eq:pmm_eem}, \ref{eq:pmm_eet} and \ref{eq:pmm_emt} for two fixed energies 2.5 GeV and 5 GeV as functions of $\delta$.
We have already observed before that for the disappearance channel, it is the higher energy range that contributes more. 
To understand $\Delta \pmm$, we will thus refer to the more relevant bottom row of Fig.~\ref{fig:deg_pmm}. 
It is clear from the figure (first and second column) that the two terms for $\Delta \pmm$ act in the same direction 
for $\eem$ (thereby increasing the overall $|\Delta \pmm|$), but show opposite behaviour for $\eet$, leading to an overall very small $\Delta \pmm$ through cancellation in the latter case. 
Looking back at Eqns.~\ref{eq:pmm_eem} and \ref{eq:pmm_eet}, we note that both term I and II are roughly 
proportional to $\cos\delta$. 
But due to the presence of a relative sign in the 
coefficient of $A \eeta$ in the second term, this behaves in almost opposite direction of the first.%
~\footnote{$s_{13} \sim 0.15, \qquad As_{23}\eema \sim Ac_{23}\eeta \sim 0.03, \qquad D(\eema) \sim D(\eeta) \lesssim 0.15\\  \alpha s_{13} s_{2(12)}  \sim 0.004, \qquad \alpha s_{2(12)} \sim 0.027, \qquad \theta(\eema) \sim \theta(\eeta) \lesssim 10^{o}$\\
Thus in the first term of the Eq.~\ref{eq:pmm_eem} and Eq.~\ref{eq:pmm_eet},  $\cos\delta$ part is dominating and in the second term, $\theta(\varepsilon_{\alpha\beta})$ is very small,- making the overall $\Delta \pmm$ approximately proportional to $\cos\delta$ for ease of understanding.} 
This has an interesting consequence that $\Delta \pmm (\eet)$ is significantly small at higher energies unlike 
$\Delta \pmm (\eem)$. 
This is also manifestly evident from our simulation (Fig.~\ref{fig:eps_deg_en}: bottom row, first and second columns). 
Additionally, in Fig.~\ref{fig:pmm_eps_deg_del} (bottom row, first and second column)
 we have also observed the appearance of two red peaks around $\pm \pi/2$ for $\eem$ and mostly reddish region (implying very small $\Delta \pmm$) in presence of $\eet$.
  
 Finally, we see from Eq.~\ref{eq:pmm_emt} and the corresponding third column of Fig.~\ref{fig:deg_pmm} that 
 $\Delta \pmm (\emt)$ is independent of the CP phase $\delta$ and its value is quite significant (except around 2.5 GeV) compared to that in presence of $\eem$ or $\eet$.
 This corroborates the 
 observations in Fig.~\ref{fig:eps_deg_en} (bottom row, third column) and Fig.~\ref{fig:pmm_eps_deg_del} (third column).

 \section{SI-NSI difference at the level of event rates in the context of DUNE}
  \label{sec:event}

% 

%%%%%%%%%%%%%%%%%%%%%%%%%%%%%%%%%%%%%%%%%%%%%%%%%%%%%%%%%%%%%%%%%%%%%%%%%%% events plot
\begin{figure*}[htb]
\centering
\includegraphics[width=\textwidth]
{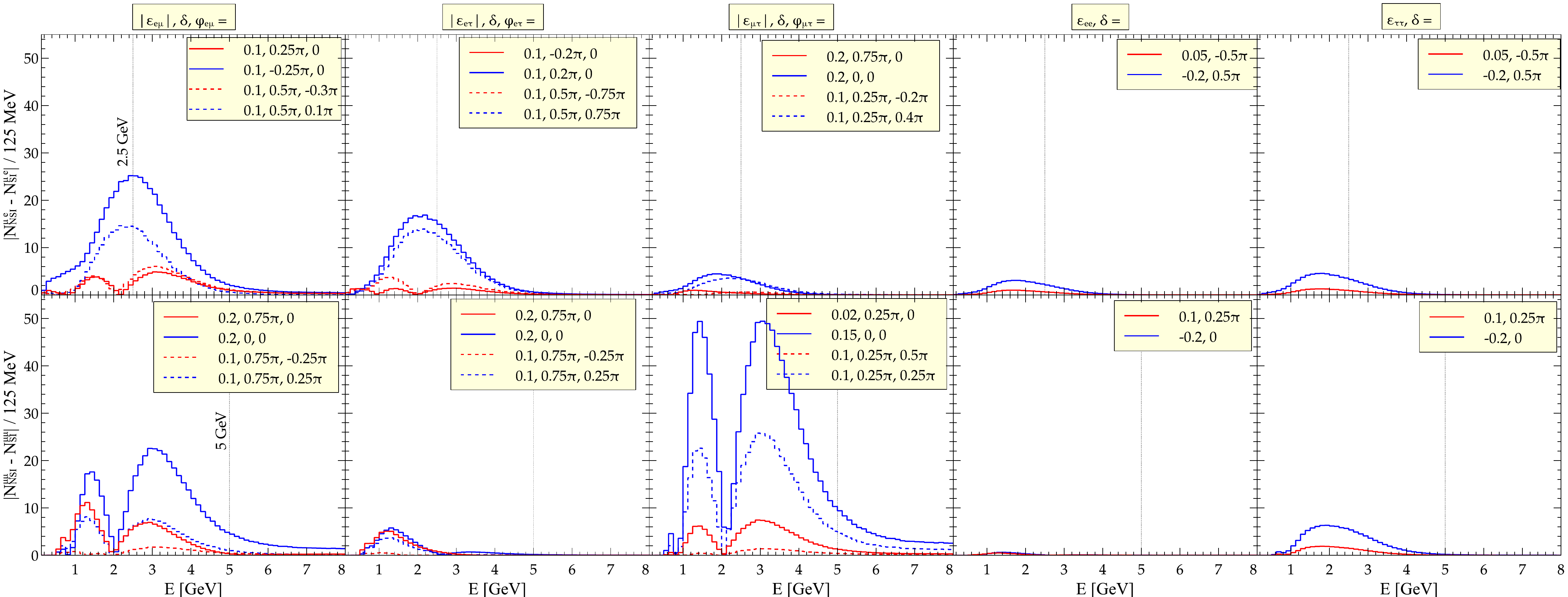}
\caption{\footnotesize{ 
SI-NSI difference at the level of event rates for $\nu_\mu \to \nu_e$ channel (top row) and $\nu_\mu \to \nu_\mu$ channel (bottom row)  and for different NSI parameters. The LE  flux has been used.}} 
\label{fig:10}
\end{figure*}

 In order to illustrate the SI-NSI degeneracy at the level of event rates, we can define the following quantity 
  \bea
  \Delta N_{\alpha\beta} (E) = {N_{\alpha\beta}^{\rm{NSI}} (E) - N_{\alpha\beta}^{\rm{SI}} (E)}
  \label{eq:event}
  \eea
  where $N_{\alpha\beta}$ stands for the  number of events for  $\nu_{\alpha} \to \nu_{\beta}$.
 The results are shown in Fig.~\ref{fig:10}. The top row depicts the event difference in case of 
 $\nu_\mu \to \nu_e$ channel and the bottom row shows the event difference 
 in case of $\nu_\mu \to \nu_\mu$ channel. We have picked four  choices of the parameters as indicated 
 in the figure. 
These choices are guided by our observations in Sec.~\ref{sec:prob}. 
%{\hlpm{For $\nu_\mu \to \nu_e$ channel (top row), the solid red and solid blue curves depict choices from Fig.~\ref{fig:pme_eps_deg_del} while the dashed ones are from Fig.~\ref{fig:pme_phi_deg_del}. }}
The red curves correspond to the almost degenerate case while blue curves correspond to regions away from degeneracy. The vertical grey line is showing the location of $2.5$ GeV ($5$ GeV) in the top panel (bottom panel). 
If we use a given beam tune (say, the standard LE beam tune), the characteristic shape of the event difference spectrum is similar to the original event spectrum in case of no degeneracy (see the blue solid and dashed curves). When we choose the parameters corresponding to degenerate solutions, the spectrum shape of the event difference is completely altered (see the red curves). 
In the latter case, one can note that the SI-NSI degeneracy manifests itself in the form of a dip near the energy value of 2.5 GeV at which first oscillation maximum occurs for $\nu_\mu \to \nu_e$ channel. 
 
Some of the crucial features that can be seen from Fig.~\ref{fig:10}  are :
\begin{itemize}
\item $\varepsilon_{e\mu}$ and $\varepsilon_{e\tau}$ have the largest impact in case of $\nu_\mu\to \nu_e$ channel  (top row of Fig.~\ref{fig:10}). 

\item $\varepsilon_{\mu\tau}$ has the largest impact in case of $\nu_\mu\to \nu_\mu$ channel (bottom row of Fig.~\ref{fig:10}).

\end{itemize} 
% 
%%%%%%%%%%%%%%%%%%%%%%%%%%%%% 

%----------------------------------------------------------

\section*{Acknowledgements} 
We acknowledge Mary Bishai for valuable discussions and for providing us the medium energy flux files used for analysis in the present work. It is a pleasure to thank Raj Gandhi for useful discussions and critical comments on the manuscript. PM  would like to thank the AHEP group at IFIC, University of Valencia for the warm hospitality during the finishing stages of this work.
 We acknowledge the use of HRI cluster facility to carry out computations in this work. 
 The work of MM is supported by the Spanish grants SEV-2014-0398, FPA2017-85216-P (AEI/FEDER, UE),
PROMETEO/2018/165 (Generalitat Valenciana) and the Spanish Red Consolider MultiDark FPA2-17-90566-REDC.
SR is supported by INFOSYS scholarship for senior students.
The work of PM is supported by the Indian funding from University Grants Commission under the second phase of University with Potential of Excellence (UPE II) at JNU and Department of Science and Technology under DST-PURSE at JNU. This work has received partial funding from the European Union's Horizon 2020 research and innovation programme under the Marie Skodowska-Curie grant agreement No 690575 and 674896.

\bibliographystyle{apsrev}
\bibliography{referencesnsi}

\begin{thebibliography}{83}
\expandafter\ifx\csname natexlab\endcsname\relax\def\natexlab#1{#1}\fi
\expandafter\ifx\csname bibnamefont\endcsname\relax
  \def\bibnamefont#1{#1}\fi
\expandafter\ifx\csname bibfnamefont\endcsname\relax
  \def\bibfnamefont#1{#1}\fi
\expandafter\ifx\csname citenamefont\endcsname\relax
  \def\citenamefont#1{#1}\fi
\expandafter\ifx\csname url\endcsname\relax
  \def\url#1{\texttt{#1}}\fi
\expandafter\ifx\csname urlprefix\endcsname\relax\def\urlprefix{URL }\fi
\providecommand{\bibinfo}[2]{#2}
\providecommand{\eprint}[2][]{\url{#2}}

\bibitem[{\citenamefont{Wolfenstein}(1978)}]{Wolfenstein:1977ue}
\bibinfo{author}{\bibfnamefont{L.}~\bibnamefont{Wolfenstein}},
  \bibinfo{journal}{Phys. Rev.} \textbf{\bibinfo{volume}{D17}},
  \bibinfo{pages}{2369} (\bibinfo{year}{1978}).

\bibitem[{\citenamefont{Beringer et~al.}(2012)}]{Beringer:1900zz}
\bibinfo{author}{\bibfnamefont{J.}~\bibnamefont{Beringer}} \bibnamefont{et~al.}
  (\bibinfo{collaboration}{Particle Data Group}), \bibinfo{journal}{Phys. Rev.}
  \textbf{\bibinfo{volume}{D86}}, \bibinfo{pages}{010001}
  (\bibinfo{year}{2012}).

\bibitem[{\citenamefont{Capozzi et~al.}(2017)\citenamefont{Capozzi,
  Di~Valentino, Lisi, Marrone, Melchiorri, and Palazzo}}]{Capozzi:2017ipn}
\bibinfo{author}{\bibfnamefont{F.}~\bibnamefont{Capozzi}},
  \bibinfo{author}{\bibfnamefont{E.}~\bibnamefont{Di~Valentino}},
  \bibinfo{author}{\bibfnamefont{E.}~\bibnamefont{Lisi}},
  \bibinfo{author}{\bibfnamefont{A.}~\bibnamefont{Marrone}},
  \bibinfo{author}{\bibfnamefont{A.}~\bibnamefont{Melchiorri}},
  \bibnamefont{and} \bibinfo{author}{\bibfnamefont{A.}~\bibnamefont{Palazzo}},
  \bibinfo{journal}{Phys. Rev.} \textbf{\bibinfo{volume}{D95}},
  \bibinfo{pages}{096014} (\bibinfo{year}{2017}), \eprint{1703.04471}.

\bibitem[{\citenamefont{de~Salas et~al.}(2018)\citenamefont{de~Salas, Forero,
  Ternes, Tortola, and Valle}}]{deSalas:2017kay}
\bibinfo{author}{\bibfnamefont{P.~F.} \bibnamefont{de~Salas}},
  \bibinfo{author}{\bibfnamefont{D.~V.} \bibnamefont{Forero}},
  \bibinfo{author}{\bibfnamefont{C.~A.} \bibnamefont{Ternes}},
  \bibinfo{author}{\bibfnamefont{M.}~\bibnamefont{Tortola}}, \bibnamefont{and}
  \bibinfo{author}{\bibfnamefont{J.~W.~F.} \bibnamefont{Valle}},
  \bibinfo{journal}{Phys. Lett.} \textbf{\bibinfo{volume}{B782}},
  \bibinfo{pages}{633} (\bibinfo{year}{2018}), \eprint{1708.01186}.

\bibitem[{\citenamefont{Valencia-Globalfit}(2018)}]{globalfit}
\bibinfo{author}{\bibnamefont{Valencia-Globalfit}},
  \bibinfo{howpublished}{\url{http://globalfit.astroparticles.es/}}
  (\bibinfo{year}{2018}).

\bibitem[{\citenamefont{Esteban et~al.}(2019)\citenamefont{Esteban,
  Gonzalez-Garcia, Hernandez-Cabezudo, Maltoni, and Schwetz}}]{Esteban:2018azc}
\bibinfo{author}{\bibfnamefont{I.}~\bibnamefont{Esteban}},
  \bibinfo{author}{\bibfnamefont{M.~C.} \bibnamefont{Gonzalez-Garcia}},
  \bibinfo{author}{\bibfnamefont{A.}~\bibnamefont{Hernandez-Cabezudo}},
  \bibinfo{author}{\bibfnamefont{M.}~\bibnamefont{Maltoni}}, \bibnamefont{and}
  \bibinfo{author}{\bibfnamefont{T.}~\bibnamefont{Schwetz}},
  \bibinfo{journal}{JHEP} \textbf{\bibinfo{volume}{01}}, \bibinfo{pages}{106}
  (\bibinfo{year}{2019}), \eprint{1811.05487}.

\bibitem[{\citenamefont{NuFIT-Globalfit}(2018)}]{nufit}
\bibinfo{author}{\bibnamefont{NuFIT-Globalfit}},
  \bibinfo{howpublished}{\url{http://www.nu-fit.org}} (\bibinfo{year}{2018}).

\bibitem[{\citenamefont{Barger et~al.}(2002)\citenamefont{Barger, Marfatia, and
  Whisnant}}]{Barger:2001yr}
\bibinfo{author}{\bibfnamefont{V.}~\bibnamefont{Barger}},
  \bibinfo{author}{\bibfnamefont{D.}~\bibnamefont{Marfatia}}, \bibnamefont{and}
  \bibinfo{author}{\bibfnamefont{K.}~\bibnamefont{Whisnant}},
  \bibinfo{journal}{Phys. Rev.} \textbf{\bibinfo{volume}{D65}},
  \bibinfo{pages}{073023} (\bibinfo{year}{2002}), \eprint{hep-ph/0112119}.

\bibitem[{\citenamefont{Gandhi et~al.}(2006)\citenamefont{Gandhi, Ghoshal,
  Goswami, Mehta, and Sankar}}]{Gandhi:2004bj}
\bibinfo{author}{\bibfnamefont{R.}~\bibnamefont{Gandhi}},
  \bibinfo{author}{\bibfnamefont{P.}~\bibnamefont{Ghoshal}},
  \bibinfo{author}{\bibfnamefont{S.}~\bibnamefont{Goswami}},
  \bibinfo{author}{\bibfnamefont{P.}~\bibnamefont{Mehta}}, \bibnamefont{and}
  \bibinfo{author}{\bibfnamefont{S.~U.} \bibnamefont{Sankar}},
  \bibinfo{journal}{Phys.Rev.} \textbf{\bibinfo{volume}{D73}},
  \bibinfo{pages}{053001} (\bibinfo{year}{2006}), \eprint{hep-ph/0411252}.

\bibitem[{\citenamefont{Huber et~al.}(2005{\natexlab{a}})\citenamefont{Huber,
  Maltoni, and Schwetz}}]{Huber:2005ep}
\bibinfo{author}{\bibfnamefont{P.}~\bibnamefont{Huber}},
  \bibinfo{author}{\bibfnamefont{M.}~\bibnamefont{Maltoni}}, \bibnamefont{and}
  \bibinfo{author}{\bibfnamefont{T.}~\bibnamefont{Schwetz}},
  \bibinfo{journal}{Phys. Rev.} \textbf{\bibinfo{volume}{D71}},
  \bibinfo{pages}{053006} (\bibinfo{year}{2005}{\natexlab{a}}),
  \eprint{hep-ph/0501037}.

\bibitem[{\citenamefont{Hagiwara et~al.}(2006)\citenamefont{Hagiwara, Okamura,
  and Senda}}]{Hagiwara:2005pe}
\bibinfo{author}{\bibfnamefont{K.}~\bibnamefont{Hagiwara}},
  \bibinfo{author}{\bibfnamefont{N.}~\bibnamefont{Okamura}}, \bibnamefont{and}
  \bibinfo{author}{\bibfnamefont{K.-i.} \bibnamefont{Senda}},
  \bibinfo{journal}{Phys. Lett.} \textbf{\bibinfo{volume}{B637}},
  \bibinfo{pages}{266} (\bibinfo{year}{2006}), \bibinfo{note}{[Erratum: Phys.
  Lett.B641,491(2006)]}, \eprint{hep-ph/0504061}.

\bibitem[{\citenamefont{Kajita et~al.}(2007)\citenamefont{Kajita, Minakata,
  Nakayama, and Nunokawa}}]{Kajita:2006bt}
\bibinfo{author}{\bibfnamefont{T.}~\bibnamefont{Kajita}},
  \bibinfo{author}{\bibfnamefont{H.}~\bibnamefont{Minakata}},
  \bibinfo{author}{\bibfnamefont{S.}~\bibnamefont{Nakayama}}, \bibnamefont{and}
  \bibinfo{author}{\bibfnamefont{H.}~\bibnamefont{Nunokawa}},
  \bibinfo{journal}{Phys. Rev.} \textbf{\bibinfo{volume}{D75}},
  \bibinfo{pages}{013006} (\bibinfo{year}{2007}), \eprint{hep-ph/0609286}.

\bibitem[{\citenamefont{Ghosh et~al.}(2016)\citenamefont{Ghosh, Ghoshal,
  Goswami, Nath, and Raut}}]{Ghosh:2015ena}
\bibinfo{author}{\bibfnamefont{M.}~\bibnamefont{Ghosh}},
  \bibinfo{author}{\bibfnamefont{P.}~\bibnamefont{Ghoshal}},
  \bibinfo{author}{\bibfnamefont{S.}~\bibnamefont{Goswami}},
  \bibinfo{author}{\bibfnamefont{N.}~\bibnamefont{Nath}}, \bibnamefont{and}
  \bibinfo{author}{\bibfnamefont{S.~K.} \bibnamefont{Raut}},
  \bibinfo{journal}{Phys. Rev.} \textbf{\bibinfo{volume}{D93}},
  \bibinfo{pages}{013013} (\bibinfo{year}{2016}), \eprint{1504.06283}.

\bibitem[{\citenamefont{Farzan and Tortola}(2018)}]{Farzan:2017xzy}
\bibinfo{author}{\bibfnamefont{Y.}~\bibnamefont{Farzan}} \bibnamefont{and}
  \bibinfo{author}{\bibfnamefont{M.}~\bibnamefont{Tortola}},
  \bibinfo{journal}{Front.in Phys.} \textbf{\bibinfo{volume}{6}},
  \bibinfo{pages}{10} (\bibinfo{year}{2018}), \eprint{1710.09360}.

\bibitem[{\citenamefont{Masud et~al.}(2016)\citenamefont{Masud, Chatterjee, and
  Mehta}}]{Masud:2015xva}
\bibinfo{author}{\bibfnamefont{M.}~\bibnamefont{Masud}},
  \bibinfo{author}{\bibfnamefont{A.}~\bibnamefont{Chatterjee}},
  \bibnamefont{and} \bibinfo{author}{\bibfnamefont{P.}~\bibnamefont{Mehta}},
  \bibinfo{journal}{J. Phys.} \textbf{\bibinfo{volume}{G43}},
  \bibinfo{pages}{095005} (\bibinfo{year}{2016}), \eprint{1510.08261}.

\bibitem[{\citenamefont{de~Gouv{\^e}a and Kelly}(2016)}]{deGouvea:2015ndi}
\bibinfo{author}{\bibfnamefont{A.}~\bibnamefont{de~Gouv{\^e}a}}
  \bibnamefont{and} \bibinfo{author}{\bibfnamefont{K.~J.} \bibnamefont{Kelly}},
  \bibinfo{journal}{Nucl. Phys.} \textbf{\bibinfo{volume}{B908}},
  \bibinfo{pages}{318} (\bibinfo{year}{2016}), \eprint{1511.05562}.

\bibitem[{\citenamefont{Coloma}(2016)}]{Coloma:2015kiu}
\bibinfo{author}{\bibfnamefont{P.}~\bibnamefont{Coloma}},
  \bibinfo{journal}{JHEP} \textbf{\bibinfo{volume}{03}}, \bibinfo{pages}{016}
  (\bibinfo{year}{2016}), \eprint{1511.06357}.

\bibitem[{\citenamefont{Liao et~al.}(2016)\citenamefont{Liao, Marfatia, and
  Whisnant}}]{Liao:2016hsa}
\bibinfo{author}{\bibfnamefont{J.}~\bibnamefont{Liao}},
  \bibinfo{author}{\bibfnamefont{D.}~\bibnamefont{Marfatia}}, \bibnamefont{and}
  \bibinfo{author}{\bibfnamefont{K.}~\bibnamefont{Whisnant}},
  \bibinfo{journal}{Phys. Rev.} \textbf{\bibinfo{volume}{D93}},
  \bibinfo{pages}{093016} (\bibinfo{year}{2016}), \eprint{1601.00927}.

\bibitem[{\citenamefont{Forero and Huber}(2016)}]{Forero:2016cmb}
\bibinfo{author}{\bibfnamefont{D.~V.} \bibnamefont{Forero}} \bibnamefont{and}
  \bibinfo{author}{\bibfnamefont{P.}~\bibnamefont{Huber}},
  \bibinfo{journal}{Phys. Rev. Lett.} \textbf{\bibinfo{volume}{117}},
  \bibinfo{pages}{031801} (\bibinfo{year}{2016}), \eprint{1601.03736}.

\bibitem[{\citenamefont{Huitu et~al.}(2016)\citenamefont{Huitu, KŠrkkŠinen,
  Maalampi, and Vihonen}}]{Huitu:2016bmb}
\bibinfo{author}{\bibfnamefont{K.}~\bibnamefont{Huitu}},
  \bibinfo{author}{\bibfnamefont{T.~J.} \bibnamefont{KŠrkkŠinen}},
  \bibinfo{author}{\bibfnamefont{J.}~\bibnamefont{Maalampi}}, \bibnamefont{and}
  \bibinfo{author}{\bibfnamefont{S.}~\bibnamefont{Vihonen}},
  \bibinfo{journal}{Phys. Rev.} \textbf{\bibinfo{volume}{D93}},
  \bibinfo{pages}{053016} (\bibinfo{year}{2016}), \eprint{1601.07730}.

\bibitem[{\citenamefont{Bakhti and Farzan}(2016)}]{Bakhti:2016prn}
\bibinfo{author}{\bibfnamefont{P.}~\bibnamefont{Bakhti}} \bibnamefont{and}
  \bibinfo{author}{\bibfnamefont{Y.}~\bibnamefont{Farzan}},
  \bibinfo{journal}{JHEP} \textbf{\bibinfo{volume}{07}}, \bibinfo{pages}{109}
  (\bibinfo{year}{2016}), \eprint{1602.07099}.

\bibitem[{\citenamefont{Masud and Mehta}(2016{\natexlab{a}})}]{Masud:2016bvp}
\bibinfo{author}{\bibfnamefont{M.}~\bibnamefont{Masud}} \bibnamefont{and}
  \bibinfo{author}{\bibfnamefont{P.}~\bibnamefont{Mehta}},
  \bibinfo{journal}{Phys. Rev.} \textbf{\bibinfo{volume}{D94}},
  \bibinfo{pages}{013014} (\bibinfo{year}{2016}{\natexlab{a}}),
  \eprint{1603.01380}.

\bibitem[{\citenamefont{Soumya and Mohanta}(2016)}]{Soumya:2016enw}
\bibinfo{author}{\bibfnamefont{C.}~\bibnamefont{Soumya}} \bibnamefont{and}
  \bibinfo{author}{\bibfnamefont{R.}~\bibnamefont{Mohanta}},
  \bibinfo{journal}{Phys. Rev.} \textbf{\bibinfo{volume}{D94}},
  \bibinfo{pages}{053008} (\bibinfo{year}{2016}), \eprint{1603.02184}.

\bibitem[{\citenamefont{Rashed and Datta}(2017)}]{Rashed:2016rda}
\bibinfo{author}{\bibfnamefont{A.}~\bibnamefont{Rashed}} \bibnamefont{and}
  \bibinfo{author}{\bibfnamefont{A.}~\bibnamefont{Datta}},
  \bibinfo{journal}{Int. J. Mod. Phys.} \textbf{\bibinfo{volume}{A32}},
  \bibinfo{pages}{1750060} (\bibinfo{year}{2017}), \eprint{1603.09031}.

\bibitem[{\citenamefont{Coloma and Schwetz}(2016)}]{Coloma:2016gei}
\bibinfo{author}{\bibfnamefont{P.}~\bibnamefont{Coloma}} \bibnamefont{and}
  \bibinfo{author}{\bibfnamefont{T.}~\bibnamefont{Schwetz}},
  \bibinfo{journal}{Phys. Rev.} \textbf{\bibinfo{volume}{D94}},
  \bibinfo{pages}{055005} (\bibinfo{year}{2016}), \bibinfo{note}{[Erratum:
  Phys. Rev.D95,no.7,079903(2017)]}, \eprint{1604.05772}.

\bibitem[{\citenamefont{Babu et~al.}(2016)\citenamefont{Babu, McKay, Mocioiu,
  and Pakvasa}}]{Babu:2016fdt}
\bibinfo{author}{\bibfnamefont{K.~S.} \bibnamefont{Babu}},
  \bibinfo{author}{\bibfnamefont{D.~W.} \bibnamefont{McKay}},
  \bibinfo{author}{\bibfnamefont{I.}~\bibnamefont{Mocioiu}}, \bibnamefont{and}
  \bibinfo{author}{\bibfnamefont{S.}~\bibnamefont{Pakvasa}},
  \bibinfo{journal}{Phys. Rev.} \textbf{\bibinfo{volume}{D93}},
  \bibinfo{pages}{113019} (\bibinfo{year}{2016}), \eprint{1605.03625}.

\bibitem[{\citenamefont{de~Gouvêa and Kelly}(2016)}]{deGouvea:2016pom}
\bibinfo{author}{\bibfnamefont{A.}~\bibnamefont{de~Gouvêa}} \bibnamefont{and}
  \bibinfo{author}{\bibfnamefont{K.~J.} \bibnamefont{Kelly}}
  (\bibinfo{year}{2016}), \eprint{1605.09376}.

\bibitem[{\citenamefont{Masud and Mehta}(2016{\natexlab{b}})}]{Masud:2016gcl}
\bibinfo{author}{\bibfnamefont{M.}~\bibnamefont{Masud}} \bibnamefont{and}
  \bibinfo{author}{\bibfnamefont{P.}~\bibnamefont{Mehta}},
  \bibinfo{journal}{Phys. Rev.} \textbf{\bibinfo{volume}{D94}},
  \bibinfo{pages}{053007} (\bibinfo{year}{2016}{\natexlab{b}}),
  \eprint{1606.05662}.

\bibitem[{\citenamefont{Blennow et~al.}(2016)\citenamefont{Blennow, Choubey,
  Ohlsson, Pramanik, and Raut}}]{Blennow:2016etl}
\bibinfo{author}{\bibfnamefont{M.}~\bibnamefont{Blennow}},
  \bibinfo{author}{\bibfnamefont{S.}~\bibnamefont{Choubey}},
  \bibinfo{author}{\bibfnamefont{T.}~\bibnamefont{Ohlsson}},
  \bibinfo{author}{\bibfnamefont{D.}~\bibnamefont{Pramanik}}, \bibnamefont{and}
  \bibinfo{author}{\bibfnamefont{S.~K.} \bibnamefont{Raut}},
  \bibinfo{journal}{JHEP} \textbf{\bibinfo{volume}{08}}, \bibinfo{pages}{090}
  (\bibinfo{year}{2016}), \eprint{1606.08851}.

\bibitem[{\citenamefont{Agarwalla et~al.}(2016)\citenamefont{Agarwalla,
  Chatterjee, and Palazzo}}]{Agarwalla:2016fkh}
\bibinfo{author}{\bibfnamefont{S.~K.} \bibnamefont{Agarwalla}},
  \bibinfo{author}{\bibfnamefont{S.~S.} \bibnamefont{Chatterjee}},
  \bibnamefont{and} \bibinfo{author}{\bibfnamefont{A.}~\bibnamefont{Palazzo}},
  \bibinfo{journal}{Phys. Lett.} \textbf{\bibinfo{volume}{B762}},
  \bibinfo{pages}{64} (\bibinfo{year}{2016}), \eprint{1607.01745}.

\bibitem[{\citenamefont{Ge and Smirnov}(2016)}]{Ge:2016dlx}
\bibinfo{author}{\bibfnamefont{S.-F.} \bibnamefont{Ge}} \bibnamefont{and}
  \bibinfo{author}{\bibfnamefont{A.~{\relax Yu}.} \bibnamefont{Smirnov}},
  \bibinfo{journal}{JHEP} \textbf{\bibinfo{volume}{10}}, \bibinfo{pages}{138}
  (\bibinfo{year}{2016}), \eprint{1607.08513}.

\bibitem[{\citenamefont{Forero and Huang}(2017)}]{Forero:2016ghr}
\bibinfo{author}{\bibfnamefont{D.~V.} \bibnamefont{Forero}} \bibnamefont{and}
  \bibinfo{author}{\bibfnamefont{W.-C.} \bibnamefont{Huang}},
  \bibinfo{journal}{JHEP} \textbf{\bibinfo{volume}{03}}, \bibinfo{pages}{018}
  (\bibinfo{year}{2017}), \eprint{1608.04719}.

\bibitem[{\citenamefont{Liao et~al.}(2017{\natexlab{a}})\citenamefont{Liao,
  Marfatia, and Whisnant}}]{Liao:2016bgf}
\bibinfo{author}{\bibfnamefont{J.}~\bibnamefont{Liao}},
  \bibinfo{author}{\bibfnamefont{D.}~\bibnamefont{Marfatia}}, \bibnamefont{and}
  \bibinfo{author}{\bibfnamefont{K.}~\bibnamefont{Whisnant}},
  \bibinfo{journal}{Phys. Lett.} \textbf{\bibinfo{volume}{B767}},
  \bibinfo{pages}{350} (\bibinfo{year}{2017}{\natexlab{a}}),
  \eprint{1609.01786}.

\bibitem[{\citenamefont{Blennow et~al.}(2017)\citenamefont{Blennow, Coloma,
  Fernandez-Martinez, Hernandez-Garcia, and Lopez-Pavon}}]{Blennow:2016jkn}
\bibinfo{author}{\bibfnamefont{M.}~\bibnamefont{Blennow}},
  \bibinfo{author}{\bibfnamefont{P.}~\bibnamefont{Coloma}},
  \bibinfo{author}{\bibfnamefont{E.}~\bibnamefont{Fernandez-Martinez}},
  \bibinfo{author}{\bibfnamefont{J.}~\bibnamefont{Hernandez-Garcia}},
  \bibnamefont{and}
  \bibinfo{author}{\bibfnamefont{J.}~\bibnamefont{Lopez-Pavon}},
  \bibinfo{journal}{JHEP} \textbf{\bibinfo{volume}{04}}, \bibinfo{pages}{153}
  (\bibinfo{year}{2017}), \eprint{1609.08637}.

\bibitem[{\citenamefont{Fukasawa et~al.}(2017)\citenamefont{Fukasawa, Ghosh,
  and Yasuda}}]{Fukasawa:2016lew}
\bibinfo{author}{\bibfnamefont{S.}~\bibnamefont{Fukasawa}},
  \bibinfo{author}{\bibfnamefont{M.}~\bibnamefont{Ghosh}}, \bibnamefont{and}
  \bibinfo{author}{\bibfnamefont{O.}~\bibnamefont{Yasuda}},
  \bibinfo{journal}{Phys. Rev.} \textbf{\bibinfo{volume}{D95}},
  \bibinfo{pages}{055005} (\bibinfo{year}{2017}), \eprint{1611.06141}.

\bibitem[{\citenamefont{Deepthi et~al.}(2017)\citenamefont{Deepthi, Goswami,
  and Nath}}]{Deepthi:2016erc}
\bibinfo{author}{\bibfnamefont{K.~N.} \bibnamefont{Deepthi}},
  \bibinfo{author}{\bibfnamefont{S.}~\bibnamefont{Goswami}}, \bibnamefont{and}
  \bibinfo{author}{\bibfnamefont{N.}~\bibnamefont{Nath}},
  \bibinfo{journal}{Phys. Rev.} \textbf{\bibinfo{volume}{D96}},
  \bibinfo{pages}{075023} (\bibinfo{year}{2017}), \eprint{1612.00784}.

\bibitem[{\citenamefont{Liao et~al.}(2017{\natexlab{b}})\citenamefont{Liao,
  Marfatia, and Whisnant}}]{Liao:2016orc}
\bibinfo{author}{\bibfnamefont{J.}~\bibnamefont{Liao}},
  \bibinfo{author}{\bibfnamefont{D.}~\bibnamefont{Marfatia}}, \bibnamefont{and}
  \bibinfo{author}{\bibfnamefont{K.}~\bibnamefont{Whisnant}},
  \bibinfo{journal}{JHEP} \textbf{\bibinfo{volume}{01}}, \bibinfo{pages}{071}
  (\bibinfo{year}{2017}{\natexlab{b}}), \eprint{1612.01443}.

\bibitem[{\citenamefont{Soumya and Mohanta}(2017)}]{C.:2017yqh}
\bibinfo{author}{\bibfnamefont{C.}~\bibnamefont{Soumya}} \bibnamefont{and}
  \bibinfo{author}{\bibfnamefont{R.}~\bibnamefont{Mohanta}},
  \bibinfo{journal}{Eur. Phys. J.} \textbf{\bibinfo{volume}{C77}},
  \bibinfo{pages}{32} (\bibinfo{year}{2017}), \eprint{1701.00327}.

\bibitem[{\citenamefont{Rout et~al.}(2017)\citenamefont{Rout, Masud, and
  Mehta}}]{Rout:2017udo}
\bibinfo{author}{\bibfnamefont{J.}~\bibnamefont{Rout}},
  \bibinfo{author}{\bibfnamefont{M.}~\bibnamefont{Masud}}, \bibnamefont{and}
  \bibinfo{author}{\bibfnamefont{P.}~\bibnamefont{Mehta}},
  \bibinfo{journal}{Phys. Rev. D} \textbf{\bibinfo{volume}{95}},
  \bibinfo{pages}{075035} (\bibinfo{year}{2017}).

\bibitem[{\citenamefont{Ghosh and Yasuda}(2017{\natexlab{a}})}]{Ghosh:2017ged}
\bibinfo{author}{\bibfnamefont{M.}~\bibnamefont{Ghosh}} \bibnamefont{and}
  \bibinfo{author}{\bibfnamefont{O.}~\bibnamefont{Yasuda}},
  \bibinfo{journal}{Phys. Rev.} \textbf{\bibinfo{volume}{D96}},
  \bibinfo{pages}{013001} (\bibinfo{year}{2017}{\natexlab{a}}),
  \eprint{1702.06482}.

\bibitem[{\citenamefont{Kelly}(2017)}]{Kelly:2017kch}
\bibinfo{author}{\bibfnamefont{K.~J.} \bibnamefont{Kelly}},
  \bibinfo{journal}{Phys. Rev.} \textbf{\bibinfo{volume}{D95}},
  \bibinfo{pages}{115009} (\bibinfo{year}{2017}), \eprint{1703.00448}.

\bibitem[{\citenamefont{Shoemaker}(2017)}]{Shoemaker:2017lzs}
\bibinfo{author}{\bibfnamefont{I.~M.} \bibnamefont{Shoemaker}},
  \bibinfo{journal}{Phys. Rev.} \textbf{\bibinfo{volume}{D95}},
  \bibinfo{pages}{115028} (\bibinfo{year}{2017}), \eprint{1703.05774}.

\bibitem[{\citenamefont{Ghosh and Yasuda}(2017{\natexlab{b}})}]{Ghosh:2017lim}
\bibinfo{author}{\bibfnamefont{M.}~\bibnamefont{Ghosh}} \bibnamefont{and}
  \bibinfo{author}{\bibfnamefont{O.}~\bibnamefont{Yasuda}}
  (\bibinfo{year}{2017}{\natexlab{b}}), \eprint{1709.08264}.

\bibitem[{\citenamefont{Deepthi et~al.}(2018)\citenamefont{Deepthi, Goswami,
  and Nath}}]{Deepthi:2017gxg}
\bibinfo{author}{\bibfnamefont{K.~N.} \bibnamefont{Deepthi}},
  \bibinfo{author}{\bibfnamefont{S.}~\bibnamefont{Goswami}}, \bibnamefont{and}
  \bibinfo{author}{\bibfnamefont{N.}~\bibnamefont{Nath}},
  \bibinfo{journal}{Nucl. Phys.} \textbf{\bibinfo{volume}{B936}},
  \bibinfo{pages}{91} (\bibinfo{year}{2018}), \eprint{1711.04840}.

\bibitem[{\citenamefont{Wang and Zhou}(2019)}]{Wang:2018dwk}
\bibinfo{author}{\bibfnamefont{T.}~\bibnamefont{Wang}} \bibnamefont{and}
  \bibinfo{author}{\bibfnamefont{Y.-L.} \bibnamefont{Zhou}},
  \bibinfo{journal}{Phys. Rev.} \textbf{\bibinfo{volume}{D99}},
  \bibinfo{pages}{035039} (\bibinfo{year}{2019}), \eprint{1801.05656}.

\bibitem[{\citenamefont{Choudhury et~al.}(2018)\citenamefont{Choudhury, Ghosh,
  and Niyogi}}]{Choudhury:2018xsm}
\bibinfo{author}{\bibfnamefont{D.}~\bibnamefont{Choudhury}},
  \bibinfo{author}{\bibfnamefont{K.}~\bibnamefont{Ghosh}}, \bibnamefont{and}
  \bibinfo{author}{\bibfnamefont{S.}~\bibnamefont{Niyogi}}
  (\bibinfo{year}{2018}), \eprint{1801.01513}.

\bibitem[{\citenamefont{Falkowski et~al.}(2018)\citenamefont{Falkowski,
  Grilli~di Cortona, and Tabrizi}}]{Falkowski:2018dmy}
\bibinfo{author}{\bibfnamefont{A.}~\bibnamefont{Falkowski}},
  \bibinfo{author}{\bibfnamefont{G.}~\bibnamefont{Grilli~di Cortona}},
  \bibnamefont{and} \bibinfo{author}{\bibfnamefont{Z.}~\bibnamefont{Tabrizi}},
  \bibinfo{journal}{JHEP} \textbf{\bibinfo{volume}{04}}, \bibinfo{pages}{101}
  (\bibinfo{year}{2018}), \eprint{1802.08296}.

\bibitem[{\citenamefont{Dey et~al.}(2018)\citenamefont{Dey, Nath, and
  Sadhukhan}}]{Dey:2018yht}
\bibinfo{author}{\bibfnamefont{U.~K.} \bibnamefont{Dey}},
  \bibinfo{author}{\bibfnamefont{N.}~\bibnamefont{Nath}}, \bibnamefont{and}
  \bibinfo{author}{\bibfnamefont{S.}~\bibnamefont{Sadhukhan}},
  \bibinfo{journal}{Phys. Rev.} \textbf{\bibinfo{volume}{D98}},
  \bibinfo{pages}{055004} (\bibinfo{year}{2018}), \eprint{1804.05808}.

\bibitem[{\citenamefont{Meloni}(2018)}]{Meloni:2018xnk}
\bibinfo{author}{\bibfnamefont{D.}~\bibnamefont{Meloni}},
  \bibinfo{journal}{JHEP} \textbf{\bibinfo{volume}{08}}, \bibinfo{pages}{028}
  (\bibinfo{year}{2018}), \eprint{1805.01747}.

\bibitem[{\citenamefont{Flores et~al.}(2018)\citenamefont{Flores, Garcés, and
  Miranda}}]{Flores:2018kwk}
\bibinfo{author}{\bibfnamefont{L.~J.} \bibnamefont{Flores}},
  \bibinfo{author}{\bibfnamefont{E.~A.} \bibnamefont{Garcés}},
  \bibnamefont{and} \bibinfo{author}{\bibfnamefont{O.~G.}
  \bibnamefont{Miranda}}, \bibinfo{journal}{Phys. Rev.}
  \textbf{\bibinfo{volume}{D98}}, \bibinfo{pages}{035030}
  (\bibinfo{year}{2018}), \eprint{1806.07951}.

\bibitem[{\citenamefont{Hyde}(2018)}]{Hyde:2018tqt}
\bibinfo{author}{\bibfnamefont{J.~M.} \bibnamefont{Hyde}}
  (\bibinfo{year}{2018}), \eprint{1806.09221}.

\bibitem[{\citenamefont{Verma and Bhardwaj}(2018)}]{Verma:2018gwi}
\bibinfo{author}{\bibfnamefont{S.}~\bibnamefont{Verma}} \bibnamefont{and}
  \bibinfo{author}{\bibfnamefont{S.}~\bibnamefont{Bhardwaj}}
  (\bibinfo{year}{2018}), \eprint{1808.04263}.

\bibitem[{\citenamefont{Chatterjee et~al.}(2018)\citenamefont{Chatterjee,
  Kamiya, Moura, and Yu}}]{Chatterjee:2018dyd}
\bibinfo{author}{\bibfnamefont{A.}~\bibnamefont{Chatterjee}},
  \bibinfo{author}{\bibfnamefont{F.}~\bibnamefont{Kamiya}},
  \bibinfo{author}{\bibfnamefont{C.~A.} \bibnamefont{Moura}}, \bibnamefont{and}
  \bibinfo{author}{\bibfnamefont{J.}~\bibnamefont{Yu}} (\bibinfo{year}{2018}),
  \eprint{1809.09313}.

\bibitem[{\citenamefont{Bischer and Rodejohann}(2019)}]{Bischer:2018zcz}
\bibinfo{author}{\bibfnamefont{I.}~\bibnamefont{Bischer}} \bibnamefont{and}
  \bibinfo{author}{\bibfnamefont{W.}~\bibnamefont{Rodejohann}},
  \bibinfo{journal}{Phys. Rev.} \textbf{\bibinfo{volume}{D99}},
  \bibinfo{pages}{036006} (\bibinfo{year}{2019}), \eprint{1810.02220}.

\bibitem[{\citenamefont{Abe et~al.}(2014)}]{Abe:2013hdq}
\bibinfo{author}{\bibfnamefont{K.}~\bibnamefont{Abe}} \bibnamefont{et~al.}
  (\bibinfo{collaboration}{T2K}), \bibinfo{journal}{Phys. Rev. Lett.}
  \textbf{\bibinfo{volume}{112}}, \bibinfo{pages}{061802}
  (\bibinfo{year}{2014}), \eprint{1311.4750}.

\bibitem[{\citenamefont{Abe et~al.}(2015)}]{Abe:2015zbg}
\bibinfo{author}{\bibfnamefont{K.}~\bibnamefont{Abe}} \bibnamefont{et~al.}
  (\bibinfo{collaboration}{Hyper-Kamiokande Proto-Collaboration}),
  \bibinfo{journal}{PTEP} \textbf{\bibinfo{volume}{2015}},
  \bibinfo{pages}{053C02} (\bibinfo{year}{2015}), \eprint{1502.05199}.

\bibitem[{\citenamefont{Abe et~al.}(2018)}]{Abe:2016ero}
\bibinfo{author}{\bibfnamefont{K.}~\bibnamefont{Abe}} \bibnamefont{et~al.}
  (\bibinfo{collaboration}{Hyper-Kamiokande}), \bibinfo{journal}{PTEP}
  \textbf{\bibinfo{volume}{2018}}, \bibinfo{pages}{063C01}
  (\bibinfo{year}{2018}), \eprint{1611.06118}.

\bibitem[{\citenamefont{Ayres et~al.}(2004)}]{Ayres:2004js}
\bibinfo{author}{\bibfnamefont{D.~S.} \bibnamefont{Ayres}} \bibnamefont{et~al.}
  (\bibinfo{collaboration}{NOvA}) (\bibinfo{year}{2004}),
  \eprint{hep-ex/0503053}.

\bibitem[{\citenamefont{Acciarri et~al.}(2015)}]{Acciarri:2015uup}
\bibinfo{author}{\bibfnamefont{R.}~\bibnamefont{Acciarri}} \bibnamefont{et~al.}
  (\bibinfo{collaboration}{DUNE}) (\bibinfo{year}{2015}), \eprint{1512.06148}.

\bibitem[{\citenamefont{Acciarri
  et~al.}(2016{\natexlab{a}})}]{Acciarri:2016ooe}
\bibinfo{author}{\bibfnamefont{R.}~\bibnamefont{Acciarri}} \bibnamefont{et~al.}
  (\bibinfo{collaboration}{DUNE}) (\bibinfo{year}{2016}{\natexlab{a}}),
  \eprint{1601.02984}.

\bibitem[{\citenamefont{Agarwalla et~al.}(2014)}]{Agarwalla:2013vyc}
\bibinfo{author}{\bibfnamefont{S.~K.} \bibnamefont{Agarwalla}}
  \bibnamefont{et~al.} (\bibinfo{collaboration}{LAGUNA-LBNO}),
  \bibinfo{journal}{JHEP} \textbf{\bibinfo{volume}{05}}, \bibinfo{pages}{094}
  (\bibinfo{year}{2014}), \eprint{1312.6520}.

\bibitem[{\citenamefont{Masud et~al.}(2019)\citenamefont{Masud, Bishai, and
  Mehta}}]{Masud:2017bcf}
\bibinfo{author}{\bibfnamefont{M.}~\bibnamefont{Masud}},
  \bibinfo{author}{\bibfnamefont{M.}~\bibnamefont{Bishai}}, \bibnamefont{and}
  \bibinfo{author}{\bibfnamefont{P.}~\bibnamefont{Mehta}},
  \bibinfo{journal}{Sci. Rep.} \textbf{\bibinfo{volume}{9}},
  \bibinfo{pages}{352} (\bibinfo{year}{2019}), \eprint{1704.08650}.

\bibitem[{\citenamefont{Biggio et~al.}(2009)\citenamefont{Biggio, Blennow, and
  Fernandez-Martinez}}]{Biggio:2009nt}
\bibinfo{author}{\bibfnamefont{C.}~\bibnamefont{Biggio}},
  \bibinfo{author}{\bibfnamefont{M.}~\bibnamefont{Blennow}}, \bibnamefont{and}
  \bibinfo{author}{\bibfnamefont{E.}~\bibnamefont{Fernandez-Martinez}},
  \bibinfo{journal}{JHEP} \textbf{\bibinfo{volume}{0908}}, \bibinfo{pages}{090}
  (\bibinfo{year}{2009}), \eprint{0907.0097}.

\bibitem[{\citenamefont{Mitsuka et~al.}(2011)}]{Mitsuka:2011ty}
\bibinfo{author}{\bibfnamefont{G.}~\bibnamefont{Mitsuka}} \bibnamefont{et~al.}
  (\bibinfo{collaboration}{Super-Kamiokande Collaboration}),
  \bibinfo{journal}{Phys.Rev.} \textbf{\bibinfo{volume}{D84}},
  \bibinfo{pages}{113008} (\bibinfo{year}{2011}), \eprint{1109.1889}.

\bibitem[{\citenamefont{Ohlsson}(2013)}]{Ohlsson:2012kf}
\bibinfo{author}{\bibfnamefont{T.}~\bibnamefont{Ohlsson}},
  \bibinfo{journal}{Rept. Prog. Phys.} \textbf{\bibinfo{volume}{76}},
  \bibinfo{pages}{044201} (\bibinfo{year}{2013}), \eprint{1209.2710}.

\bibitem[{\citenamefont{Adamson et~al.}(2013)}]{Adamson:2013ovz}
\bibinfo{author}{\bibfnamefont{P.}~\bibnamefont{Adamson}} \bibnamefont{et~al.}
  (\bibinfo{collaboration}{MINOS Collaboration}), \bibinfo{journal}{Phys.Rev.}
  \textbf{\bibinfo{volume}{D88}}, \bibinfo{pages}{072011}
  (\bibinfo{year}{2013}), \eprint{1303.5314}.

\bibitem[{\citenamefont{Kopp et~al.}(2010)\citenamefont{Kopp, Machado, and
  Parke}}]{Kopp:2010qt}
\bibinfo{author}{\bibfnamefont{J.}~\bibnamefont{Kopp}},
  \bibinfo{author}{\bibfnamefont{P.~A.} \bibnamefont{Machado}},
  \bibnamefont{and} \bibinfo{author}{\bibfnamefont{S.~J.} \bibnamefont{Parke}},
  \bibinfo{journal}{Phys.Rev.} \textbf{\bibinfo{volume}{D82}},
  \bibinfo{pages}{113002} (\bibinfo{year}{2010}), \eprint{1009.0014}.

\bibitem[{\citenamefont{Acciarri
  et~al.}(2016{\natexlab{b}})}]{Acciarri:2016crz}
\bibinfo{author}{\bibfnamefont{R.}~\bibnamefont{Acciarri}} \bibnamefont{et~al.}
  (\bibinfo{collaboration}{DUNE}) (\bibinfo{year}{2016}{\natexlab{b}}),
  \eprint{1601.05471}.

\bibitem[{\citenamefont{Abi et~al.}(2018)}]{Abi:2018dnh}
\bibinfo{author}{\bibfnamefont{B.}~\bibnamefont{Abi}} \bibnamefont{et~al.}
  (\bibinfo{collaboration}{DUNE}) (\bibinfo{year}{2018}), \eprint{1807.10334}.

\bibitem[{\citenamefont{Huber et~al.}(2005{\natexlab{b}})\citenamefont{Huber,
  Lindner, and Winter}}]{Huber:2004ka}
\bibinfo{author}{\bibfnamefont{P.}~\bibnamefont{Huber}},
  \bibinfo{author}{\bibfnamefont{M.}~\bibnamefont{Lindner}}, \bibnamefont{and}
  \bibinfo{author}{\bibfnamefont{W.}~\bibnamefont{Winter}},
  \bibinfo{journal}{Comput. Phys. Commun.} \textbf{\bibinfo{volume}{167}},
  \bibinfo{pages}{195} (\bibinfo{year}{2005}{\natexlab{b}}),
  \eprint{hep-ph/0407333}.

\bibitem[{\citenamefont{Huber et~al.}(2007)\citenamefont{Huber, Kopp, Lindner,
  Rolinec, and Winter}}]{Huber:2007ji}
\bibinfo{author}{\bibfnamefont{P.}~\bibnamefont{Huber}},
  \bibinfo{author}{\bibfnamefont{J.}~\bibnamefont{Kopp}},
  \bibinfo{author}{\bibfnamefont{M.}~\bibnamefont{Lindner}},
  \bibinfo{author}{\bibfnamefont{M.}~\bibnamefont{Rolinec}}, \bibnamefont{and}
  \bibinfo{author}{\bibfnamefont{W.}~\bibnamefont{Winter}},
  \bibinfo{journal}{Comput. Phys. Commun.} \textbf{\bibinfo{volume}{177}},
  \bibinfo{pages}{432} (\bibinfo{year}{2007}), \eprint{hep-ph/0701187}.

\bibitem[{\citenamefont{Alion et~al.}(2016)}]{Alion:2016uaj}
\bibinfo{author}{\bibfnamefont{T.}~\bibnamefont{Alion}} \bibnamefont{et~al.}
  (\bibinfo{collaboration}{DUNE}) (\bibinfo{year}{2016}), \eprint{1606.09550}.

\bibitem[{\citenamefont{Dziewonski and Anderson}(1981)}]{Dziewonski:1981xy}
\bibinfo{author}{\bibfnamefont{A.~M.} \bibnamefont{Dziewonski}}
  \bibnamefont{and} \bibinfo{author}{\bibfnamefont{D.~L.}
  \bibnamefont{Anderson}}, \bibinfo{journal}{Phys. Earth Planet. Interiors}
  \textbf{\bibinfo{volume}{25}}, \bibinfo{pages}{297} (\bibinfo{year}{1981}).

\bibitem[{\citenamefont{Adams et~al.}(2013)}]{2013arXiv1307.7335L}
\bibinfo{author}{\bibfnamefont{C.}~\bibnamefont{Adams}} \bibnamefont{et~al.}
  (\bibinfo{collaboration}{LBNE Collaboration}), \bibinfo{journal}{ArXiv
  e-prints}  (\bibinfo{year}{2013}), \eprint{1307.7335}.

\bibitem[{\citenamefont{Agostinelli et~al.}(2003)}]{Agostinelli:2002hh}
\bibinfo{author}{\bibfnamefont{S.}~\bibnamefont{Agostinelli}}
  \bibnamefont{et~al.} (\bibinfo{collaboration}{GEANT4}),
  \bibinfo{journal}{Nucl. Instrum. Meth.} \textbf{\bibinfo{volume}{A506}},
  \bibinfo{pages}{250} (\bibinfo{year}{2003}).

\bibitem[{\citenamefont{Allison et~al.}(2006)}]{Allison:2006ve}
\bibinfo{author}{\bibfnamefont{J.}~\bibnamefont{Allison}} \bibnamefont{et~al.},
  \bibinfo{journal}{IEEE Trans. Nucl. Sci.} \textbf{\bibinfo{volume}{53}},
  \bibinfo{pages}{270} (\bibinfo{year}{2006}).

\bibitem[{\citenamefont{Kopp}(2008)}]{Kopp:2006wp}
\bibinfo{author}{\bibfnamefont{J.}~\bibnamefont{Kopp}}, \bibinfo{journal}{Int.
  J. Mod. Phys.} \textbf{\bibinfo{volume}{C19}}, \bibinfo{pages}{523}
  (\bibinfo{year}{2008}), \eprint{physics/0610206}.

\bibitem[{\citenamefont{Kopp et~al.}(2008)\citenamefont{Kopp, Lindner, Ota, and
  Sato}}]{Kopp:2007ne}
\bibinfo{author}{\bibfnamefont{J.}~\bibnamefont{Kopp}},
  \bibinfo{author}{\bibfnamefont{M.}~\bibnamefont{Lindner}},
  \bibinfo{author}{\bibfnamefont{T.}~\bibnamefont{Ota}}, \bibnamefont{and}
  \bibinfo{author}{\bibfnamefont{J.}~\bibnamefont{Sato}},
  \bibinfo{journal}{Phys. Rev.} \textbf{\bibinfo{volume}{D77}},
  \bibinfo{pages}{013007} (\bibinfo{year}{2008}), \eprint{0708.0152}.

\bibitem[{\citenamefont{Kikuchi et~al.}(2009)\citenamefont{Kikuchi, Minakata,
  and Uchinami}}]{Kikuchi:2008vq}
\bibinfo{author}{\bibfnamefont{T.}~\bibnamefont{Kikuchi}},
  \bibinfo{author}{\bibfnamefont{H.}~\bibnamefont{Minakata}}, \bibnamefont{and}
  \bibinfo{author}{\bibfnamefont{S.}~\bibnamefont{Uchinami}},
  \bibinfo{journal}{JHEP} \textbf{\bibinfo{volume}{0903}}, \bibinfo{pages}{114}
  (\bibinfo{year}{2009}), \eprint{0809.3312}.

\bibitem[{\citenamefont{Asano and Minakata}(2011)}]{Asano:2011nj}
\bibinfo{author}{\bibfnamefont{K.}~\bibnamefont{Asano}} \bibnamefont{and}
  \bibinfo{author}{\bibfnamefont{H.}~\bibnamefont{Minakata}},
  \bibinfo{journal}{JHEP} \textbf{\bibinfo{volume}{1106}}, \bibinfo{pages}{022}
  (\bibinfo{year}{2011}), \eprint{1103.4387}.

\bibitem[{\citenamefont{Esteban et~al.}(2018)\citenamefont{Esteban,
  Gonzalez-Garcia, Maltoni, Martinez-Soler, and Salvado}}]{Esteban:2018ppq}
\bibinfo{author}{\bibfnamefont{I.}~\bibnamefont{Esteban}},
  \bibinfo{author}{\bibfnamefont{M.~C.} \bibnamefont{Gonzalez-Garcia}},
  \bibinfo{author}{\bibfnamefont{M.}~\bibnamefont{Maltoni}},
  \bibinfo{author}{\bibfnamefont{I.}~\bibnamefont{Martinez-Soler}},
  \bibnamefont{and} \bibinfo{author}{\bibfnamefont{J.}~\bibnamefont{Salvado}},
  \bibinfo{journal}{JHEP} \textbf{\bibinfo{volume}{08}}, \bibinfo{pages}{180}
  (\bibinfo{year}{2018}), \eprint{1805.04530}.

\bibitem[{\citenamefont{Miranda et~al.}(2006)\citenamefont{Miranda, Tortola,
  and Valle}}]{Miranda:2004nb}
\bibinfo{author}{\bibfnamefont{O.~G.} \bibnamefont{Miranda}},
  \bibinfo{author}{\bibfnamefont{M.~A.} \bibnamefont{Tortola}},
  \bibnamefont{and} \bibinfo{author}{\bibfnamefont{J.~W.~F.}
  \bibnamefont{Valle}}, \bibinfo{journal}{JHEP} \textbf{\bibinfo{volume}{10}},
  \bibinfo{pages}{008} (\bibinfo{year}{2006}), \eprint{hep-ph/0406280}.

\bibitem[{\citenamefont{Barenboim et~al.}(2019)\citenamefont{Barenboim, Masud,
  Ternes, and Tórtola}}]{Barenboim:2018ctx}
\bibinfo{author}{\bibfnamefont{G.}~\bibnamefont{Barenboim}},
  \bibinfo{author}{\bibfnamefont{M.}~\bibnamefont{Masud}},
  \bibinfo{author}{\bibfnamefont{C.~A.} \bibnamefont{Ternes}},
  \bibnamefont{and} \bibinfo{author}{\bibfnamefont{M.}~\bibnamefont{Tórtola}},
  \bibinfo{journal}{Phys. Lett.} \textbf{\bibinfo{volume}{B788}},
  \bibinfo{pages}{308} (\bibinfo{year}{2019}), \eprint{1805.11094}.

\end{thebibliography}

\end{document}